\documentclass[aip,jcp,reprint]{revtex4-1}
\pdfoutput=1
\usepackage{amsmath,amsfonts,amssymb,bm,graphicx,hyperref}
\usepackage{color}

\newcommand{\ee}{\mathrm{e}}

\newcommand{\Weff}{W_{\rm eff}}
\newcommand{\Wapp}{W_{\rm app}}

\newcommand{\rlj}[1]{{#1}}

\begin{document}

\title{Coarse-grained depletion potentials for anisotropic colloids: application to lock-and-key systems}
\author{Clement Law}
\author{Douglas J. Ashton}
\author{Nigel B. Wilding}
\author{Robert L. Jack}
\affiliation{Department of Physics, University of Bath, Bath BA2 7AY, United Kingdom}

\begin{abstract}
When a colloid is mixed with a depletant such as a non-adsorbing polymer, one observes attractive effective interactions between the colloidal particles.  If these particles are anisotropic, analysis of these effective interactions is challenging in general.  We present a method for inference of approximate (coarse-grained) effective interaction potentials between such anisotropic particles.  Using the example of indented (lock-and-key) colloids, we show how numerical solutions can be used to integrate out the (hard sphere) depletant, leading to a depletion potential that accurately characterises the effective interactions.  The accuracy of the method is based on matching of contributions to the second virial coefficient of the colloids.  The simplest version of our method yields a piecewise-constant effective potential; we also show how this scheme can be generalised to other functional forms, where appropriate.
\end{abstract}

\maketitle

\section{Introduction}

Colloidal systems of spherical particles have been studied
extensively, and support a wide range of behavior, including solid and
fluid phases in one-component systems, and liquid-vapor phase
transitions in colloid-polymer mixtures~\cite{Poon02}.  The behavior of
\emph{anisotropic} colloids is even richer, including liquid crystals~\cite{Onsager1949,Mukhija:2011aa}, 
exotic crystalline phases~\cite{Rossi:2015aa,Chen:2011eu}, and liquids with unusual structure~\cite{Ashton2015-porous,Ruzicka:2011ud}.
New colloidal synthesis methods have stimulated recent work in this area, with a range
of anisotropic particles now available, including indented (lock-and-key) particles~\cite{Sacanna:2010ys,Sacanna:2013kq},
fused spheres~\cite{Sacanna:2013aa,Kraft12}, ellipsoids~\cite{Dugyala:2013aa} and superballs~\cite{Rossi:2015aa,Rossi:2011qd}.
When such anisotropic particles are mixed with a non-adsorbing polymer, one
finds depletion forces between colloids that depend strongly on their
orientations \cite{Florea:2014aa, Karas:2016aa, Anders:2014la}, which can lead to self-assembly of complex structures.

The microscopic mechanism for depletion forces is well-understood~\cite{Asakura1954,Lekkerkerker:2011} --
mixing colloids with a non-adsorbing depletant leads to unbalanced
osmotic pressures on the colloids, resulting in an effective
attraction.  Depletion forces between spherical colloids can 
be characterised theoretically by integrating out the depletant~\cite{Likos:2001fk,Asakura1954,Dijkstra1999,Lekkerkerker:2011} -- whether this is easy or difficult
depends on the type of depletant particles, but quantitative
theoretical predictions can be made, at least for two body
effective interactions between the colloids\cite{Ashton:2011kx}. Three and higher body
interactions are usually harder to calculate but are
expected to be negligible in comparison to two-body forces if the
ratio of depletant size to colloid size is small\cite{Ashton2014-jcp}. By contrast,
accurate characterisation of two body
effective interactions between anisotropic particles is difficult in general. For instance in the
case of uniaxial particles (whose orientations can be described by a
single unit vector) the effective potential is a function of four
variables -- such functions may not be easy to infer or parameterise
theoretically. Particles of lower symmetry require even greater number
of variables.

In this article, we introduce a general strategy for developing
approximate (coarse-grained) interaction potential between anisotropic
particles, and we apply it to a system of indented (lock-and-key)
colloids~\cite{Sacanna:2010ys,Sacanna:2013kq}.  
\rlj{Such systems have been studied quite extensively in theory and simulation~\cite{Odriozola:2008uq,Marechal:2010fk,Odriozola2013,Ahmed:zr,Melendez2015,Villegas2016,Chang:2015aa}: despite their simplicity, they exhibit strong directional bonds, which can lead to rich phenomenology, both for packing~\cite{Ahmed:zr,Ashton2013} and phase behavior~\cite{Ashton2015-porous,Ashton2015-wall}.}

Our general coarse-graining method is designed to yield piecewise-constant
interaction potentials that match the binding free energies for the
different regimes in which anisotropic colloids can associate with
each other.  
\rlj{For systems of spherical particles with short-ranged interactions, the extended law of corresponding states~\cite{Noro2000}, means that matching these free energies leads to coarse-grained models that are very effective in reproducing systems' phase behaviour.  For anisotropic particles with strong directional binding, Wertheim's theory~\cite{Wertheim:1984zr} indicates that these free energies again control the behaviour of the system, as found for lock-and-key colloids in Ref.~\onlinecite{Ashton2013}.}
These free energies are characterised in terms of the
second virial coefficient of the coarse-grained system, so for
spherical colloids, the simplest version of the method would yield a
square-well attraction between the colloids, with a second virial
coefficient chosen to match the fully-interacting system.   For anisotropic particles, one arrives at
a more complex effective interaction, but the physical motivation is
similar, so one can hope that the coarse-grained model will match the
full system at a similar level of accuracy.  
\rlj{Hence, our method, which is tailored towards colloidal systems with hard cores and short-ranged interactions, differs from methods used in molecular or polymeric systems~\cite{Noid2008,Prapotnik2007,Likos:2001fk,Shelley2001}.}

The form of the paper is as follows: Sec.~\ref{sec:model} describes
our model and Sec.~\ref{sec:theory} describes the general theory that
we use to develop a coarse-grained effective interaction.  In
Sec.~\ref{sec:lock-key} we describe the relatively simple case of an
effective interaction between an indented colloid (a lock) and a hard
sphere (a key).  In Sec.~\ref{sec:lock-lock} we discuss the effective
interaction between two lock particles, which depends in a complex way
on the relative orientations of the two particles.
Sec.~\ref{sec:simp} addresses the relationship between our approach
here and a simplified version of this effective
potential that was used in Ref.~\onlinecite{Ashton2015-porous}. Our conclusions are summarized in
Sec.~\ref{sec:conc}.

\section{Model: lock-and-key colloids}
\label{sec:model}

\begin{figure}
\includegraphics[width=7cm]{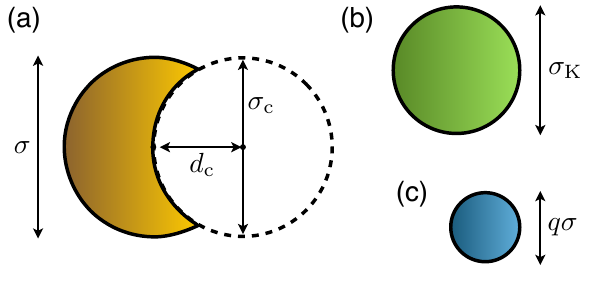}
\caption{Illustration of the different hard particles considered in this work.  (a) Indented colloid (lock) particle, defined by considering a sphere of diameter $\sigma$ and cutting away its intersection with a second sphere of diameter $\sigma_{\rm c}$.  (b) A spherical colloid (key) particle with diameter $\sigma_{\rm K}$, comparable with $\sigma$.  (c) Smaller depletant particle of diameter $q\sigma$: in this work we take $q=0.1$ so the depletant is significantly smaller than the colloidal particles.}
\label{fig:model}
\end{figure}

Our model system is based on the experimental system of Sacanna and co-workers~\cite{Sacanna:2010ys}: it was introduced in Ref.~\onlinecite{Ashton2013} and further studied in Refs.~\onlinecite{Ashton2015-porous,Ashton2015-wall}.  Similar model systems have also been studied in theory and simulation~\cite{Odriozola:2008uq,Marechal:2010fk,Odriozola2013,Ahmed:zr,Melendez2015,Villegas2016,Chang:2015aa}.  The model consists of hard particles of different sizes and shapes, as shown in Fig.~\ref{fig:model}.  To define the anisotropic particles, consider a hard spherical particle of diameter $\sigma$, in which we make a concave indentation by cutting away its intersection with a second sphere of diameter $\sigma_c$.  The distance between the centers of the original sphere and the cutting sphere is $d_c$.  The orientation of an indented colloid is described by a unit vector $\bm{n}$ that points from the center of the original sphere towards the center of the cutting sphere.  These indented particles  interact with a depletant consisting of smaller hard spheres of diameter $q\sigma$.  In some cases, we also mix these two components with additional hard spheres of diameter $\sigma_{\rm K}$.  To make contact with Refs.~\onlinecite{Ashton2013,Ashton2015-porous,Ashton2015-wall}, note that if $\sigma_{\rm c}=\sigma$ then the depth of the indentation (measured from the lip) is $h=(\sigma - d_{\rm c})/2$, so specifying the depth $h$ is equivalent to specifying the shape parameter $d_{\rm c}$.

We refer to the indented particles as \emph{lock particles}, since the spherical keys fit within the indentation, leading to \emph{lock-and-key binding}.  Compared with the interaction between spherical particles, this binding is strong, due to the complementary shapes of the lock and key particles~\cite{Kinoshita:2002kx,Konig2008,Odriozola:2008uq,Sacanna:2010ys,Anders:2014la}.  We refer to both the indented particles and the keys as \emph{colloidal particles}, to distinguish them from the depletant.  We have in mind that both species of colloidal particles have comparable sizes, while the depletant particles are considerably smaller. (In this work we take $\sigma_{\rm K}=\sigma$ and $q=0.1$ throughout.)  It is therefore useful to \emph{integrate out} the depletant degrees of freedom, to arrive at a coarse-grained system in which only the colloids survive, and the effect of the depletant is captured via a two-body effective interaction~\cite{Dijkstra1999}.  This interaction depends on the chemical potential of the depletant, which we describe in terms of its (reservoir) volume fraction $\eta$.

Our numerical method for integrating out the depletant involves
explicit simulations of a pair of colloids in a depletant fluid. To
the extent that such simulations are feasible, it is applicable to any
type of depletant fluid and is thus quite general. We note that one
can avoid explicit simulation of the depletant fluid if one assumes
that the depletant is `ideal', in which case the depletion potential
can be estimated via a numerical integration scheme. This is the
approach taken in Ref.~\onlinecite{Villegas2016}, which, in a spirit
similar to the present work, makes estimates of how the binding free
energy of lock and key colloids depends on their geometry and relative
orientations.

We have performed Monte Carlo simulations of interacting lock and key
colloids with a hard sphere depletant, and separately, colloids that
interact with each other through an effective interaction that is
designed to mimic the full colloid-depletant mixture.  In all cases we
use the geometrical cluster algorithm~\cite{Dress1995,Liu2004} (GCA)
to move the particles, following the same methods as in
Refs.~\onlinecite{Ashton2013,Ashton2015-porous}.

\section{Theory}
\label{sec:theory}

\newcommand{\rhobar}{\overline{\rho}}
\newcommand{\kB}{k_{\rm B}}

Before describing results for indented colloids, we present our general method for inferring (from simulation data) the effective interactions between anisotropic colloids.  We begin with a brief review of the situation for isotropic (spherical) particles.
In this case, the effective interaction can be defined in terms of the radial distribution function, in the dilute limit.  Given a large system of colloidal particles interacting with a depletant at (reservoir) volume fraction $\eta$, one defines
%
%For spherical colloids, a standard method of inferring the effective interaction is to simulate a system containing two colloids interacting with the depletant, and to measure the radial distribution function for the colloids, 
\begin{equation}
g_\eta(r) = \frac{\rho_\eta^{(2)}(\bm{R},\bm{R}')}{\rhobar^2}
\end{equation}
where translational invariance means that the right-hand side depends only on $r=|\bm{R}-\bm{R}'|$, we have introduced the mean colloid density $\rhobar$, and the two-body density 
\begin{equation}
\rho_\eta^{(2)}(\bm{R},\bm{R}') = \langle \rho(\bm{R}) \rho(\bm{R}') \rangle_\eta - \rhobar\delta(\bm{R}-\bm{R}').
\end{equation}
Angle brackets $\langle \cdot \rangle_\eta$ indicate equilibrium averages in the colloid-depletant mixture, with depletant volume fraction $\eta$.  Given these definitions, the (dimensionless) effective potential between the colloids can be defined as 
\begin{equation}
W_{\rm eff}^\eta(r) = \lim_{\rhobar\to0} \left[ - \log \frac{g_\eta(r)}{g_0(r)} \right]
\label{equ:weff-sph}
\end{equation}
for all $r$ where $g_0(r)>0$, and $W_{\rm eff}(r)=0$ otherwise.  Here $g_0(r)=g_{\eta=0}(r)$ is the radial distribution function in the absence of depletant.  For hard spherical colloids of diameter $\sigma$, we have that $g_0(r)\to\Theta(r-\sigma)$ as $\rhobar\to0$, so the denominator in (\ref{equ:weff-sph}) is not required, but we include it for later convenience.
%The function $g(r)$ can be calculated efficiently from a histogram of particle separations, based on the simulation of two colloids with depletant.

For anisotropic colloids, there is a corresponding two-body density
$\rho_\eta^{(2)}(\bm{R},\Omega,\bm{R}',\Omega')$ which depends on the positions $\bm{R}$ and orientations $\Omega$ of both colloids.
The corresponding effective potential is (by analogy with (\ref{equ:weff-sph}))
\begin{equation}
W_{\rm eff}^\eta(\bm{R},\Omega,\bm{R}',\Omega') = - \lim_{\rhobar\to0} \left[ \log \frac{\rho_\eta^{(2)}(\bm{R},\Omega,\bm{R}',\Omega')}{\rho_0^{(2)}(\bm{R},\Omega,\bm{R}',\Omega')} \right]
\label{equ:weff-aniso}
\end{equation}

\begin{figure}
\includegraphics[width=8cm]{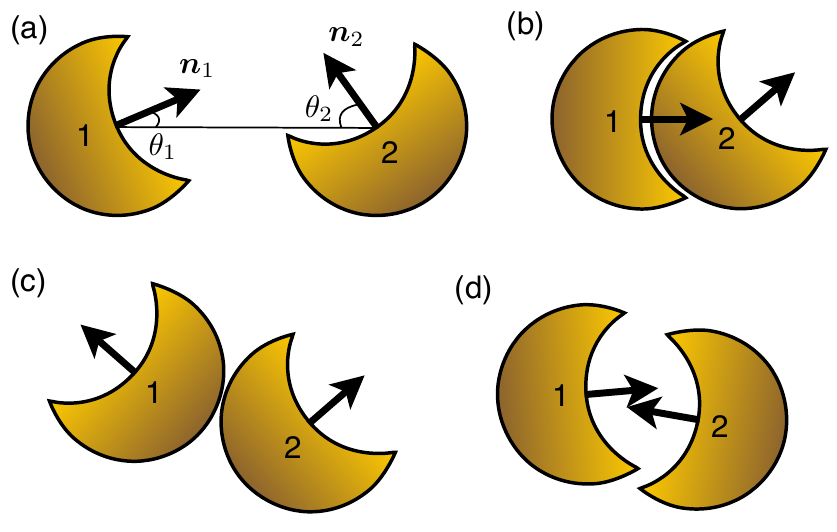}
\caption{(a) Co-ordinate system describing the relative orientation of two indented colloids. We define $\theta_{1,2}$ as the angles between the particles' orientation vectors $\bm{n}_{1,2}$ and the interparticle displacement vector. [The interparticle vector is defined between the geometric centres of the locks: here we show the case $d_{\rm c}=\sigma/2$, for which the geometrical centre of a lock lies on its concave surface.   We define $\theta_{\rm R}=\min(\theta_1,\theta_2)$ and $\theta_{\rm I}=\max(\theta_1,\theta_2)$.  The orientation vectors $\bm{n}_{1,2}$ are not in general co-planar with the interparticle vector so in order to describe the relative position and orientation of the two particles, we must also specify the angle $\phi=\cos^{-1}(\bm{n}_1\cdot\bm{n}_2)$.  All angles take values in the range $[0,\pi]$.  Interchanging the particle labels 1 and 2 leaves the angles $\theta_{\rm R},\theta_{\rm I},\phi$ invariant so any effective potential that depends on only these angles and the particle separation is automatically independent of particle labelling. (b) The specific (lock-and-key) binding regime is associated with small values of $\theta_{\rm R}$ and $\phi$ (in this case $\theta_{\rm R}=\theta_1\approx0$), and large values of $\theta_{\rm I}$   (c) The non-specific (back-to-back) regime is associated with large values of $\theta_{\rm R}$, in which case the interaction strength also depends weakly on $\theta_{\rm I},\phi$. (d) The \emph{mouth-to-mouth} regime is associated with small values of $\theta_{\rm R},\theta_{\rm I}$ and large values of $\phi$.}
\label{fig:lock-lock-coord}
\end{figure}

In contrast to the spherical case where the two-body density depends only on the distance between the particles, this two-body density depends on more than one variable.  For example, if the orientation of each colloid can be described in terms of a single orientation vector (as for the indented colloids considered here), then $W_{\rm eff}^\eta$ depends on the distance between the particles and on three angular co-ordinates that describe the relative orientation of the two colloids (see Fig.~\ref{fig:lock-lock-coord}).  This makes estimation of $W_{\rm eff}^\eta$ much more challenging for anisotropic colloids, because while $g(r)$ can be inferred from simulation data via a simple one-dimensional histogram, the direct generalisation of that method to anisotropic particles would require assembly of a four-dimensional histogram.  \rlj{
For a one-dimensional histogram, one might expect to represent the effective potential accurately using a histogram with around 100 bins. To obtain a four-dimensional histogram at similar accuracy, one would require $100^4$ bins, and assembly of such a histogram would require a data set with at least $100$ times as many samples as there are bins in the histogram.  One easily sees that this method quickly becomes unfeasible.  Moreoever,
it does not provide a simple or intuitive representation of the effective interaction.}

The method that we now present shows how the complexity of these high-dimensional distributions can be reduced by an appropriate choice of co-ordinate system, leading to a parameterisation of the effective potential.  These procedures require physical insight into the physics of the interacting system, but we show that accurate results are still available even if the complicated four-dimensional function $W_{\rm eff}$ is simplified very considerably.

\subsection{Second virial coefficients}

The key to the accuracy of our scheme is that we develop an approximate effective potential which matches precisely the second virial coefficient associated with the true effective potential.  In the absence of a depletant, we describe the interactions between colloidal particles via a two-body potential $v_0(\bm{R},\Omega,\bm{R}',\Omega')$.  We then define a second virial coefficient associated with the effective interactions among the colloids, which is
\begin{equation}
B_2^\eta = \frac12 \int [ 1 - {\rm e}^{-\beta v_{\rm eff}(\bm{R},\Omega,\bm{R}',\Omega')}  ] \mathrm{d}\bm{R}' \mathrm{d}\Omega'
\label{equ:b2}
\end{equation}
where $\beta v_{\rm eff} = \beta v_0 + W_{\rm eff}$, and $\beta=1/\kB T$ is the inverse temperature.  The right hand side of (\ref{equ:b2}) is independent of $(\bm{R},\Omega)$ since the system is translationally and rotationally invariant.  For later convenience, we define the orientational integral to be normalised such that $\int\mathrm{d}\Omega=1$ so if the orientation of the particle can be described by a single unit vector $\bm{n}$ then $\mathrm{d}\Omega = \mathrm{d}^2\bm{n}/(4\pi)$.

Now imagine fixing the position and orientation $\bm{R},\Omega$ of the first colloidal particle, and decomposing the domain of the integral in (\ref{equ:b2}) into several regions -- each region will correspond to a particular set of positions and orientations of a second particle.  For example, for the lock-shaped colloids shown in Fig.~\ref{fig:lock-lock-coord}, one such region will involve the two particles bonded in the ``back-to-back'' binding mode (Fig.~\ref{fig:lock-lock-coord}c).  The contribution of region $X$ to the second virial coefficient is
\begin{equation}
B_2^\eta(X) = \frac12 \int_X [ 1 - {\rm e}^{-\beta v_{\rm eff}(\bm{R},\Omega,\bm{R}',\Omega')} ] \mathrm{d}\bm{R}' \mathrm{d}\Omega' .
\label{equ:B2X}
\end{equation}

Our aim in this work is to define an approximate parameterisation $W_{\rm app}$ of the effective potential so that for each relevant region $X$, the integral $B_2(X)$ evaluated with the approximated potential matches the value $B_2^\eta(X)$ obtained with the depletant in place.  That is, we define 
\begin{equation}
B_2^{\rm app}(X) = \frac12 \int_X [ 1 - {\rm e}^{-\beta v_{\rm app}(\bm{R},\Omega,\bm{R}',\Omega')}]\mathrm{d}\bm{R}' \mathrm{d}\Omega' ,
\label{equ:B2app}
\end{equation} 
with $\beta v_{\rm app}=\beta v_0 + W_{\rm app}$ and we choose $W_{\rm app}$ such that for a specific set of regions $X$, we have $B_2^{\rm app}(X) = B_2^\eta(X)$.
We choose $W_{\rm app}$ according to this criterion instead of (for example) matching the \emph{values} of $W_{\rm eff}$ and $W_{\rm app}$, since $B_2(X)$ determines the probability that two particles bind together with a relative orientation $X$, and this is the most important quantity for the physical properties of the coarse-grained system.  This approach is also useful in other settings, for example in understanding the phase behaviour of systems with short-ranged interactions\cite{Noro2000,Vliegenthart2000}, or the application to anisotropic particles of Wertheim's theory of associating fluids\cite{Wertheim:1984zr,Ashton2013}.  The second virial coefficients for different binding regimes are also related to equilibrium constants associated with binding/unbinding\cite{Melendez2015,Villegas2016} [in the simplest case, one has an equilibrium constant for binding in regime $X$ which is $K_X\approx -B_2(X)$, where the approximate equality is accurate when the effective interactions are strong ($\ee^{-\beta v_{\rm eff}}\gg 1$).]

\subsection{Estimation of piecewise linear effective interactions}

Within this scheme, the simplest way to define an approximate
effective potential is to choose a set of regions $X_1,X_2,\dots$, and
take $W_{\rm app}$ to be a piecewise constant function, with a
different (constant) value in each region.  (This approach is used,
for example, when describing the attraction between spherical colloids
by a square-well potential.)  To this end, it is useful to consider a
system of just two colloidal particles in the presence of a depletant.
Integrating out the depletant yields $Z_2^\eta = (V/2) \int {\rm
  e}^{-\beta v_{\rm eff}(\bm{R},\Omega,\bm{R}',\Omega')}
\mathrm{d}\bm{R}' \mathrm{d}\Omega'$: standard results from
liquid-state theory ensure that the effective potential $W_{\rm eff}$
that appears in this integral (via $v_{\rm eff}$) is the same as that
defined in (\ref{equ:weff-aniso}). From the definition of $Z_2^\eta$, one immediately sees that
$Z_2^\eta = V [(V/2) - B_2^\eta]$.

It follows that if a large set of simulation data samples the configuration space of this system, the fraction of data points for which the relative colloid co-ordinates are in region $X$ will be
\begin{align}
P_2^\eta(X) & = \frac{V}{2Z_2^\eta} \int_X {\rm e}^{-\beta v_{\rm eff}(\bm{R},\Omega,\bm{R}',\Omega')} \mathrm{d}\bm{R}' \mathrm{d}\Omega'
\nonumber \\ & = \frac{V(X) - 2B_2^\eta(X)}{V-2B_2^\eta}
\label{equ:P2B2}
\end{align}
where $V(X) =  \int_X  \mathrm{d}\bm{R}' \mathrm{d}\Omega'$ is the volume of region $X$.  

From the definition of $B_2^{\rm app}$ in (\ref{equ:B2app}) and using the fact that $W_{\rm app}$ is constant within region $X$, we obtain $B_2^{\rm app}(X)  = \frac12 [ V(X) - \ee^{-W_{\rm app}(X)} \int_X \ee^{-\beta v_0} \mathrm{d}\bm{R}' \mathrm{d}\Omega']$.  
Choosing the value of $W_{\rm app}(X)$ such that 
$B_2^{\rm app}(X) = B_2^\eta(X)$, we obtain
\begin{align}
\ee^{-W_{\rm app}(X)} &= \frac{V(X) - 2B_2^\eta(X)}{\int_X \ee^{-\beta v_0} \mathrm{d}\bm{R}' \mathrm{d}\Omega'}
%\nonumber \\ & = \frac{V(X) - 2B_2^{\rm app}(X)}{Z^0_2 V \cdot P^0_2(X)}
\nonumber \\ & = \frac{V(X) - 2B_2^{\rm app}(X)}{V(X) - 2B_2^0(X)}
\end{align}
where the notation $B_2^0$ indicates $B_2^{\eta=0}$, so the second equality follows from (\ref{equ:b2}).
Hence we can use (\ref{equ:P2B2}) to write
\begin{align}
W_{\rm app}(X) %&= \frac{V(X) - 2B_2^{\rm app}(X)}{\int_X \ee^{-\beta v_0} \mathrm{d}\bm{R}' \mathrm{d}\Omega'}
%\nonumber \\ & = \frac{V(X) - 2B_2^{\rm app}(X)}{Z^0_2 V \cdot P^0_2(X)}
%\nonumber \\ & 
= -\log \left[ \frac{P_2^\eta(X)}{P_2^0(X)} \cdot \frac{1-(2B_2^\eta/V)}{1 - (2B_2^0/V)}\right] .
\label{equ:Wapp-P}
\end{align}

Given simulation data for two colloids interacting with depletant, and
accompanying data for two colloids alone in the simulation box, the
right hand side of (\ref{equ:Wapp-P}) can be estimated (see below).
This provides a value for $W_{\rm app}(X)$.  In this case, the choice
of the regions $X$ uniquely determines the approximate
(coarse-grained) potential -- whatever choice is made, the $B_2^{\rm
  app}(X)$ will match exactly the $B_2^\eta(X)$ evaluated for the true
depletion potential.  Similarly, the total virial coefficient $B_2$
for the coarse-grained model exactly matches that of the true
effective potential.  Computationally, the scheme is efficient because
$P_2^\eta(X)$ is a simple probability -- it avoids the requirement for
binning and making histograms from high-dimensional data sets.

We note that the factor $1-(2B_2^\eta/V)$ that appears
in~(\ref{equ:Wapp-P}) can also be obtained from the data for two
interacting particles.  Given a range $R$ that is larger than the
range of the effective interaction (but smaller than half the periodic
box), the probability that the two particles have a separation $r>R$
is easily verified to be $P_R^\eta=(V-4\pi R^3/3)/(V-2B^\eta_2)$, from
which we obtain 
\begin{equation} 
\left(1-\frac{2B_2^\eta}{V}\right) = \frac{2Z_2^\eta}{V^2} = \left( 1-\frac{4\pi R^3}{3V}\right) \frac{1}{P_R^\eta} 
\label{equ:B2-Z-PR}
\end{equation} 
Using this result, the right hand side of (\ref{equ:Wapp-P}) can be evaluated from simulation data.
Eq.~(\ref{equ:B2-Z-PR})
also allows straightforward estimation of the second virial
coefficient \cite{Ashton2014-pre,Ashton2014-jcp}.

%by calculating the probability  that the two particles have a separation $R_{12}$ between $R$ and $R+\Delta R$, where $R$ is larger than the range of the effective interaction.  As long as $R+\Delta R$ is smaller than half of the box size (to avoid problems with periodic boundaries), this probability is easily verified to be $P_R=\frac{4\pi}{3V}[(R+\Delta R)^3 - R^3]/(1-2B^\eta_2/V)$: this result also allows straightforward estimation of the second virial coefficient [cite Doug+Nigel+Bob].  
%[Alternatively, since the simulation box is finite, one may calculate the probability that $R_{12}$ is larger than $R$ and replace the numerator in $P_R$ by $1-(4\pi R^3/3V)$.]

Finally, we note that the use of a piecewise constant interaction potential is convenient because of the very simple expression (\ref{equ:Wapp-P}) that allows estimation of $W_{\rm app}$.  Such a potential is often appropriate if the resulting coarse-grained system is to be simulated by a Monte Carlo method.  In other cases, a continuous approximation to the effective potential may be required.  Matching of second virial coefficients for different regions $X$ can still be achieved in this case, but is slightly more complicated.  An example is given in Sec.~\ref{sec:lock-lock}, below.

\section{Results -- depletion potential for lock-key binding}
\label{sec:lock-key}

\begin{figure}
\includegraphics[width=8.5cm]{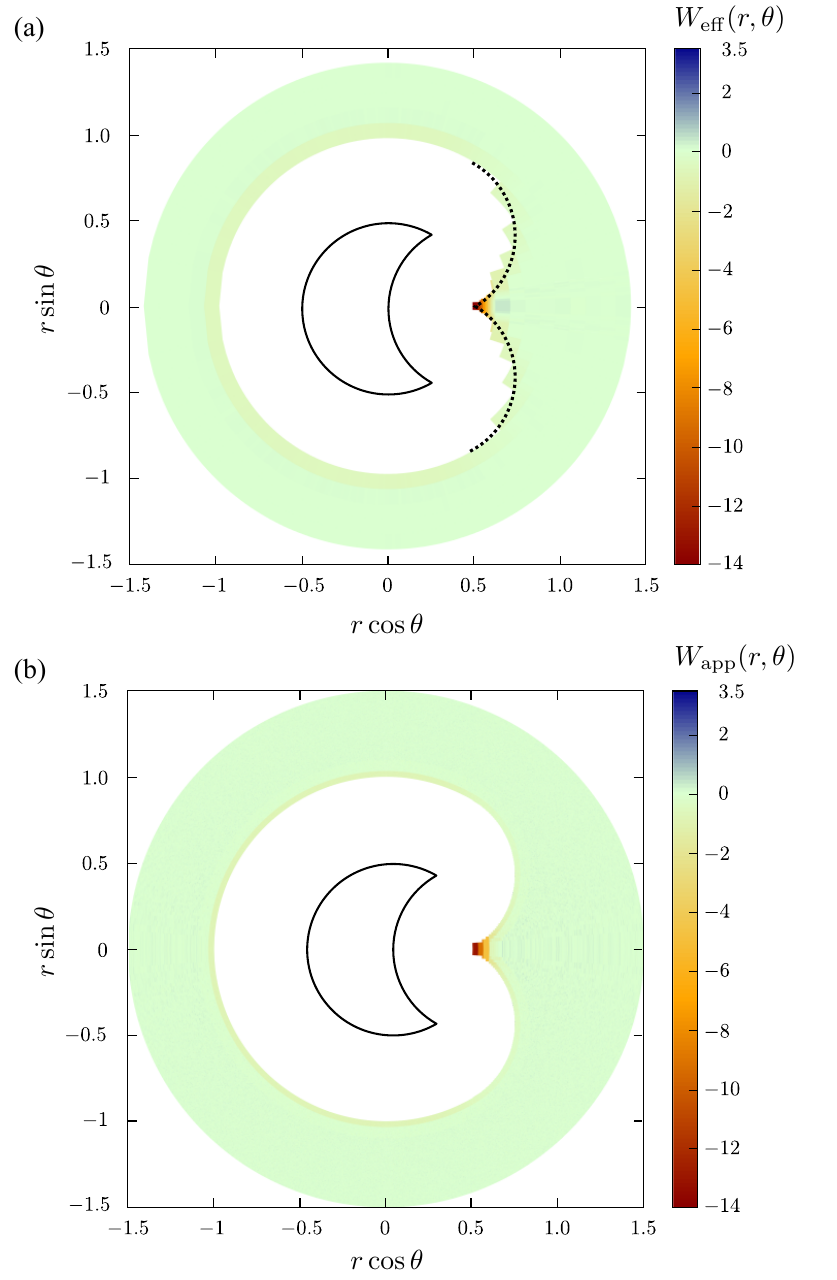}
\caption{(a) Effective potential $W_{\rm eff}(r,\cos\theta)$ 
for a lock and sphere, given a depletant with $\eta=0.08$ (b) piecewise constant approximation $W_{\rm app}$ for the same potential, inferred according to our scheme.  The potential at a given point is the free energy cost for introducing a sphere whose centre is located at that point, given that the lock particle is positioned as shown.}
\label{fig:LK-eff}
\end{figure}

\begin{figure}
\includegraphics[width=7.0cm]{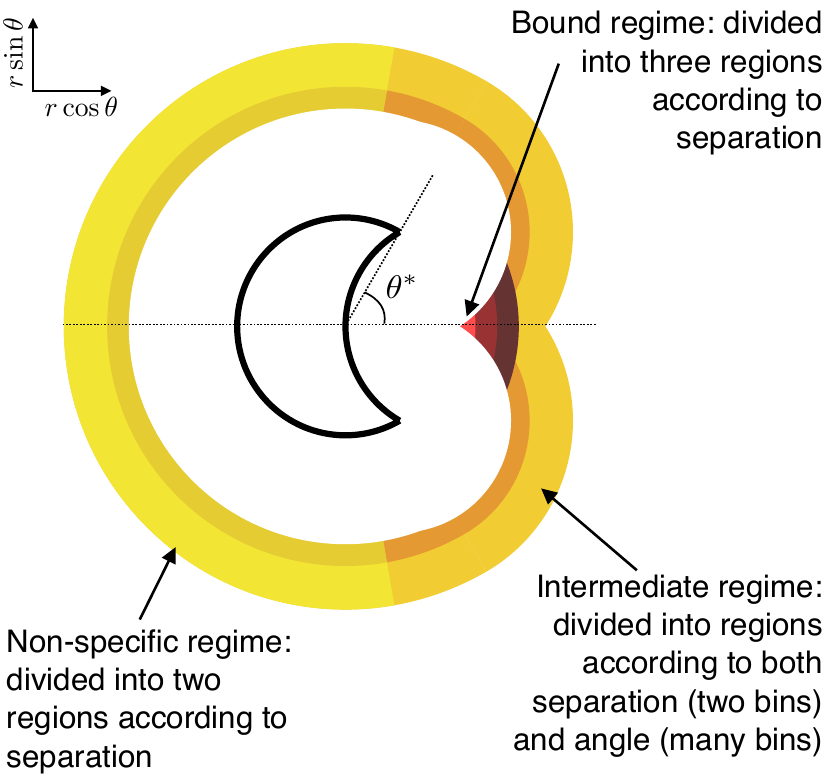}
\caption{Configuration space for a lock interacting with a sphere, partitioned into three regimes and their underlying regions.  For illustrative purposes, this figure shows a case where the range of the interaction is larger than in Fig.~\ref{fig:LK-eff}.}
\label{fig:LK-schem}
\end{figure}

So far, we have assumed that all colloids in the system are of a single species.  However, the theory presented above can easily be extended to mixtures of colloids.  In this section, we consider the effective potential between the indented (lock) particles and spherical (key) particles.  While considering mixtures might appear complicated, this situation is in fact rather simple because the effective potential depends on just two co-ordinates\cite{Odriozola:2008uq,Villegas2016}.  Let the positions of one lock and one key be $\bm{R}_{\rm L}$ and $\bm{R}_{\rm K}$ and let the orientation of the lock be $\bm{n}_{\rm L}$. Then the effective potential depends only on $r=|\bm{R}_{\rm L} - \bm{R}_{\rm K}|$ and the angle $\theta$ between the lock orientation and the interparticle vector, which can be calculated from $\cos\theta = \bm{n}_{\rm L}\cdot(\bm{R}_{\rm K} - \bm{R}_{\rm L}) /r$, taking $0\leq \theta \leq \pi$.

\subsection{Perfectly-fitting lock and key}
\label{subsec:lk-exact}

We first consider a lock particle with $\sigma_{\rm c}=\sigma$ and $d_c=0.5\sigma$, and a spherical (key) particle of diameter $\sigma_{\rm K}=\sigma$, that fits exactly within the lock.  We used the GCA to simulate one lock and one key particle, interacting with depletant particles at various volume fractions.   We constructed two-dimensional histograms of the separation $r$ and angular co-ordinate $\cos\theta$.  Taking each bin of the histogram to be a region $X$ and using (\ref{equ:Wapp-P}), we arrive at a potential $\Wapp$ that accurately represents $\Weff$. (In the limit where the bin size of the histogram goes to zero, this $\Wapp$ converges exactly to $\Weff$.)  
The resulting estimate of $\Weff$ is shown in Fig.~\ref{fig:LK-eff}(a).  

We used the same simulation results to infer an approximate (coarse-grained) effective potential $\Wapp$, as we will describe shortly.  We then performed GCA simulations for one lock and one key particle, interacting by this effective potential (in the absence of the depletant).  Fig.~\ref{fig:LK-eff}(b) shows results for this coarse-grained system.  We note that Figs.~\ref{fig:LK-eff}(a,b) are both generated from GCA simulations, using the same data analysis routines -- the differences between these figures arise because one set of GCA simulations include the depletant explicitly while the other set uses a coarse-grained model of interacting colloids.
The visual agreement between $\Weff$ and $\Wapp$ is good: as noted above, the contributions of the relevant binding regimes to the second virial coefficient also match exactly.

To construct $\Wapp$, we follow the general approach described above.   We partition the two-dimensional space parameterised by $(r,\cos\theta)$ into several different regions, as illustrated in Fig.~\ref{fig:LK-schem}.  Given this partitioning, the potential $\Wapp$ follows directly.  (For separations outside the shaded regions in Fig.~\ref{fig:LK-schem}, we take $\Wapp=0$.)  We describe the various regions in turn before summarising our main results and their dependence on the depletant volume fraction $\eta$.

The range of $r,\theta$ for which we obtain simulation data is limited by the hard core repulsion of the colloids (small-$r$) and  by the finite box size (large-$r$).  For the case of a perfectly fitting key of the same size as the lock ($\sigma=\sigma_{\rm K}=\sigma_c$), and for any $d_c$, the region forbidden by hard-core repulsion is $r(\theta) < r_0(\theta)$ with
\begin{equation}
%r(\theta) < r_0(\theta).
r_0(\theta) = \begin{cases} \sigma, & \text{if}\ \theta\geq\theta^* \\ \sigma\cos(\theta^*-\theta), & \text{if}\  \theta<\theta^* \end{cases}
\label{equ:r0-overlap}
\end{equation} 
%with $r_0(\theta)=\sigma$ for $\theta>\theta^*$ and $r_0(\theta)=\sigma\cos(\theta^*-\theta)$ for $\theta<\theta^*$. 
[The angle $\theta^*$ is defined as in Fig.~\ref{fig:LK-schem} as the angular co-ordinate of the lip of the lock.  In this section we have $\sigma_{\rm c}=\sigma_{\rm K}$, and hence $\cos\theta^*=d_{\rm c}/\sigma$.]

\subsubsection{Bound regime (specific lock-and-key binding)}

The distance of closest approach between lock and key is $d_{\rm c}$ since in this case the key coincides with the cutting sphere that is used to define the lock shape.  The strongest effective interaction between lock and key occurs when $r$ is close to $d_{\rm c}$.  This is only possible for small angles $\theta$, due to the colloidal shapes.  It is therefore sufficient to define the bound (lock-and-key) regime solely in terms of $r$: we define three regions (indexed by $n=1,2,3$) based on the distance between the colloids, which are denoted by $X_{\mathrm{LK},n}$.  As shown in Fig.~\ref{fig:LK-schem}, the $n$th region includes separations $r$ satisfying 
\begin{equation}
\tfrac13(n-1)q\sigma\leq (r-d_{\rm c})<\tfrac13 nq\sigma.
\label{equ:lk-bound-ineq}
\end{equation}
Recall that $q\sigma$ is the diameter of a depletant particle, which determines the range of the depletion attraction.
Based on numerical simulations of two colloidal particles interacting with the depletant, we evaluated (\ref{equ:Wapp-P}) for each of these three regions, and for a range of depletant volume fractions $\eta$.  For $\eta=0.08$ (the case illustrated in Fig.~\ref{fig:LK-eff}) the values of the effective potential in the three regions are
$W_{\rm app}=-13.1,-9.1,-4.9$, 
% RLJ note: these numbers come from RLJ directory lockLockTidy/clemEffResultsLK/eta_0080_gca.ini
consistent with the expected strong lock-and-key binding.  (Recall $\Weff$ and $\Wapp$ are dimensionless potentials, normalised by $\kB T$, so large negative values of $W_{\rm app}$ correspond to strong attractive forces.)

\subsubsection{Non-specific regime}

Lock-and-key binding occurs when the key particle approaches the concave surface of the lock.  However, there are also significant depletion attractions when the key approaches the convex surface of the lock~\cite{Ashton2015-porous,Melendez2015,Villegas2016}: this is similar to the depletion attraction between two spheres.  To account for this effect, we define two regions (indexed by $n=1,2$) which are denoted by $X_{\mathrm{BB},n}$ and illustrated in Fig.~\ref{fig:LK-schem}.  These regions are specified by
\begin{align}
\theta & > \theta^\dag \nonumber \nonumber \\ 
r^{{\rm BB}}_{n-1} & <  (r-\sigma) < r^{{\rm BB}}_n
\end{align}
with $(r^{{\rm BB}})_n=(0,q\sigma/3,q\sigma)$ for $n=0,1,2$, and the angle $\theta^\dag$ satisfies $\cos\theta^\dag=0.1$.  We use a radial decomposition into just two regions for simplicity: the effective potential in this case resembles that between two spheres, and depends strongly on $r$ for small $r$ while the dependence for larger $r$ is weaker.  The choice of the angle $\theta^\dag$ will be discussed in the next subsection.

For $\eta=0.08$ the effective potential in these two back-to-back regions is 
$W_{\rm app}=-0.9,0.1$, showing that this potential is weaker than the lock-key binding (as expected).  However, this attraction can still be significant, particularly since the entropy (or number of configurations) compatible with this binding mode is much larger than for the lock-and-key case.

\begin{figure}
\includegraphics[width=8cm]{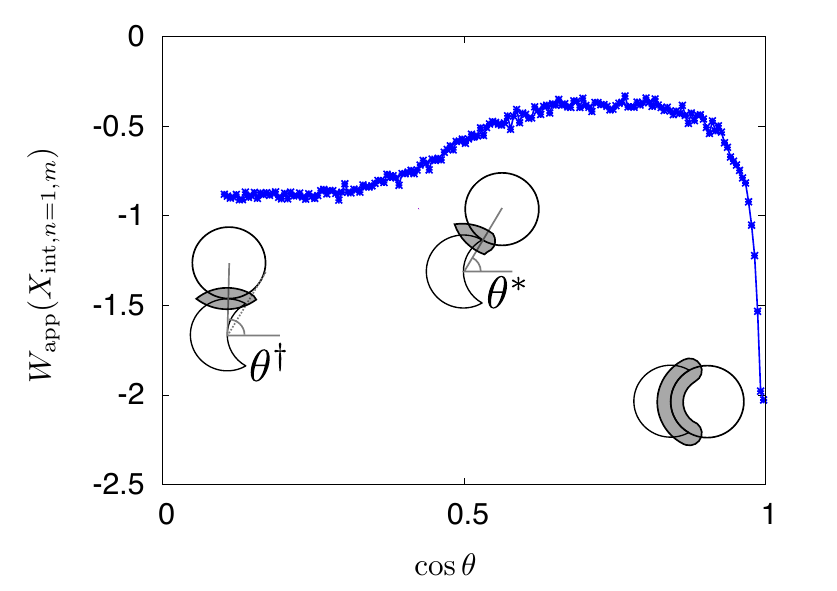}
\caption{Approximated effective potential in the intermediate regime, as a function of $\cos\theta$, for separations $r$ between $r_c(\theta)$ and  $r_c(\theta)+q\sigma/3$.  The different values of $\theta$ are illustrated by sketches, with shaded regions indicating the depletion volume: large depletion volumes correspond to strong attractive forces.  For small values of $\cos\theta$, the behaviour is similar to the non-specific binding regime.  For $\cos\theta\approx 1$, the system approaches the lock-and-key binding regime, although the strongly bound configurations are included in the bound regions, leading to the relatively weak effective potential for this regime.  In the intermediate regime, the key rolls around the lip of the lock, leading to a reduced depletion volume and therefore a reduced effective potential.}
\label{fig:LK-ang}
\end{figure}

\subsubsection{Intermediate binding regime}

The specific and non-specific binding modes considered so far tend to dominate the behaviour of this system.  In between, there is an intermediate regime, as shown in Fig.~\ref{fig:LK-schem}.  Within this regime, the effective potential depends on both the separation $r$ and the angle $\theta$.  
To capture this, we defined regions $X_{\mathrm{int},n,m}$ by 
\begin{align}
r^{{\rm BB}}_{n-1} & <  r-r_0(\theta) < r^{{\rm BB}}_n
\nonumber \\
\theta_{m-1} & < \theta < \theta_m \nonumber \nonumber 
\\ 
r &> d_c + \tfrac13 q\sigma 
\label{equ:lk-roll-ineq}
\end{align}
Here $n$ is an index associated with the particle separation ($n=1,2$) and $m=1\dots M$ is associated with the angular co-ordinate $\theta$, for which we use a larger number of bins, equally spaced in $\cos\theta$.  (In this work we have taken $M=180$ although a smaller number of regions would also be possible.)
The third inequality in (\ref{equ:lk-roll-ineq}) simply ensures that these intermediate regions do not overlap with the lock-and-key bound regions defined in (\ref{equ:lk-bound-ineq}).

For the inner region ($n=1$) we show the values of $\Wapp$ in Fig.~\ref{fig:LK-ang}, as a function of the angular co-ordinate (or equivalently the index $m$).  For large $\theta$ (or small $\cos\theta$) the behaviour is similar to the non-specifically bound regime and $\Wapp\approx -0.9$, consistent with that case.  The largest angle that falls inside the intermediate regime is $\theta_M=\theta^\dag$, and $\theta^\dag$ is chosen large enough so that the intermediate regime includes all angles for which the behaviour differs significantly from the non-specifically bound regime, hence our choice $\cos\theta^\dag=0.1$.  For small $\theta$ (or large $\cos\theta$), the system approaches the lock-and-key binding state, although there is no overlap lock-and-key bound regime.

As one passes through the intermediate regime, there is a maximum in $\Wapp$.  To explain this, we sketch in Fig.~\ref{fig:LK-ang} the depletion volume for representative configurations.  The obtain this volume, we consider for each colloid the volume that is inaccessible to the centre of a depletant particle.  As two colloids approach each other, the depletion volume is the intersection between their inaccessible volumes, which provides an estimate of the strength of the interaction~\cite{Asakura1954}. The reduction in depletion volume as the key passes through the intermediate regime explains the maximum in $\Wapp$.

\begin{figure}
\includegraphics[width=7.0cm]{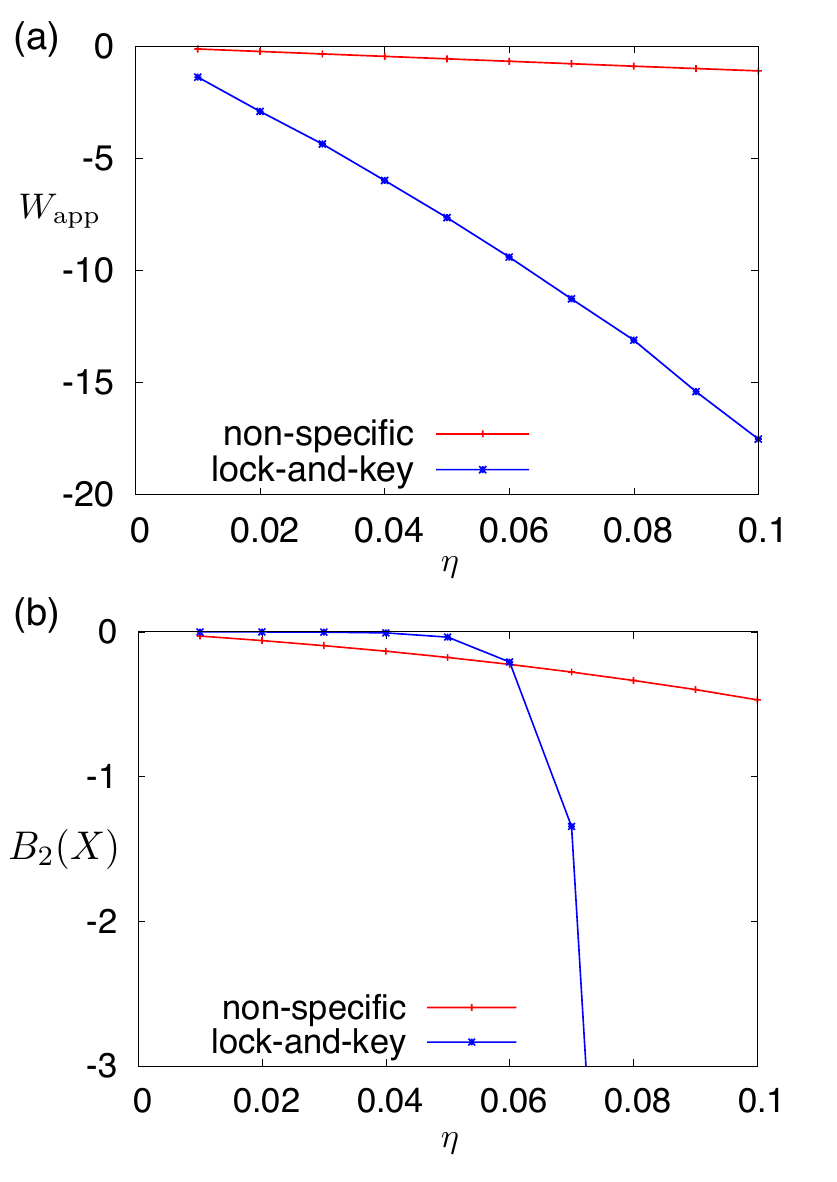}
\caption{(a) Strength of the effective interactions as a function of depletant volume fraction $\eta$.  (b) Contributions of these two binding modes to the second virial coefficient (measured in units where $\sigma=1$).  One sees that even if the effective potential is strong for lock-and-key (specific) binding, the small volume $V(X)$ of the bound regions means that the contribution of this binding mode to the second virial coefficient becomes significant only for $\eta\gtrsim 0.06$, after which it quickly becomes very strong.}
\label{fig:lk-b2}
\end{figure}

\subsubsection{Summary}

The good agreement between exact and approximated effective interactions is shown in Fig.~\ref{fig:LK-eff}.
The approximated interaction includes three parameters associated with the lock-and-key (specific) binding state, two parameters associated with back-to-back (non-specific) binding, and a lookup table for the intermediate regime.  All parameters are inferred automatically from data for the two-colloid system (with and without depletant).

Depending on the accuracy required for the effective potential in the intermediate regime, we anticipate that a considerably reduced approximate description would still be feasible and would capture the most important features of the system.  See also Sec.~\ref{sec:simp} below.

In Fig.~\ref{fig:lk-b2}(a) we plot the well depths associated with the specific and non-specific binding regimes, as a function of the depletant volume fraction $\eta$.  (These are the values of the effective potential in the innermost regions, $n=1$.)  The lock-and-key interaction strength is strong and increases strongly with $\eta$, as expected, while the non-specific binding is weaker.  However, lock-and-key binding requires localisation of the key particle in a small binding region, so the effective potential itself does not reflect the probability of binding in a given model.  In Fig.~\ref{fig:lk-b2}(b) we show the contributions of the two binding modes to the second virial coefficient (measured in units where $\sigma=1$).  The larger volume (and hence larger entropy) associated with the non-specific binding means that this binding mode is preferred for small $\eta$ (where the interactions are weak in any case), but the lock-and-key binding regime depends strongly on $\eta$ and dominates for $\eta\gtrsim 0.07$.

\subsection{Imperfectly fitting lock and key}

\begin{figure}
\includegraphics[width=8.5cm]{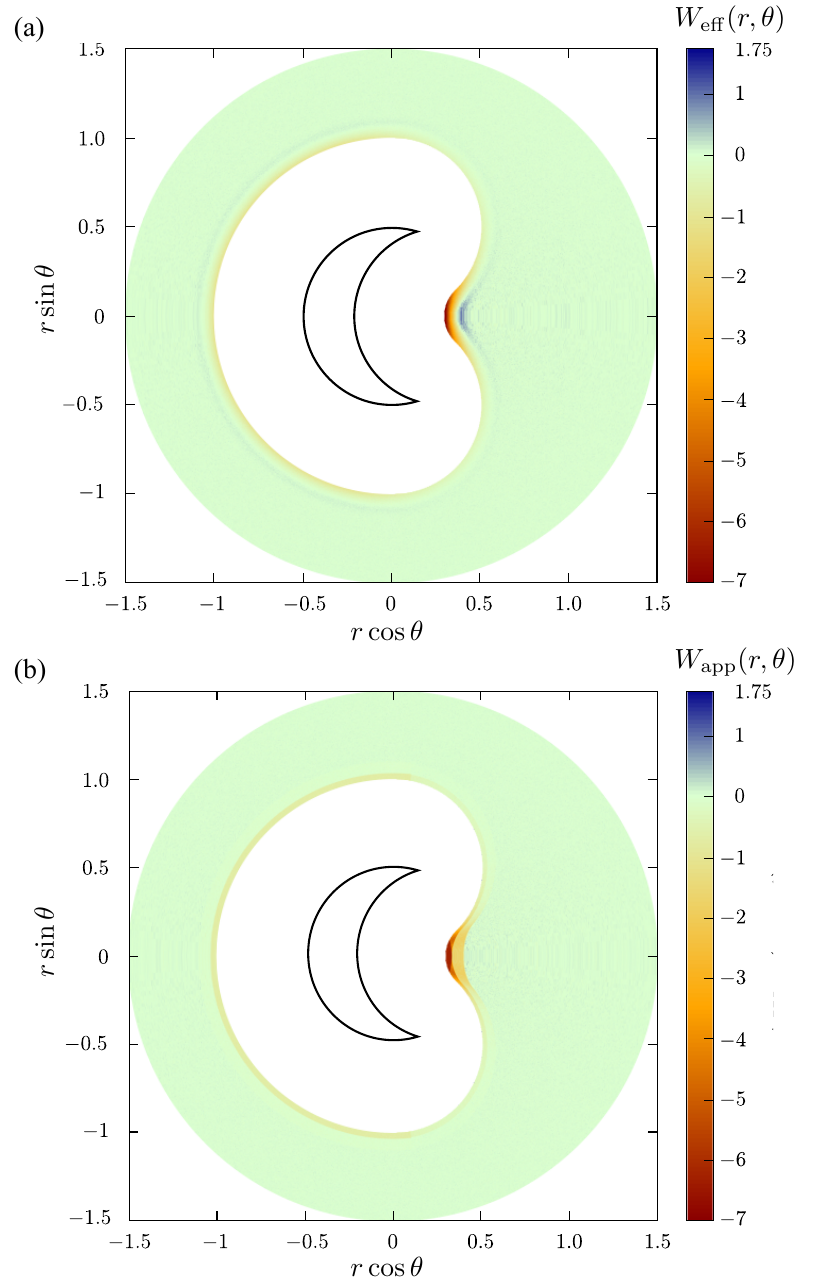}
\caption{(a) Effective potential $W_{\rm eff}$ at $\eta=0.08$ for a spherical particle that does not exactly fit the indentation in the lock particle.  (b) Piecewise constant approximation $W_{\rm app}$ to this interaction.  The repulsive region in the lock mouth is not captured: this is due to layering of the depletant particles in this region.  This effect could be captured by introducing an extra region in the approximated interaction, but we chose to ignore it, for simplicity.}
\label{fig:LK-rattle-eff}
\end{figure}

To illustrate the general applicability of this method, we now apply it to the effective interaction between a lock particle and a spherical `key' whose diameter does not precisely fit the indentation on the lock.  Specifically, we take a lock particle with $d_{\rm c}=0.5\sigma$ (as before) but $\sigma_{\rm c}=0.7\sigma$, and we keep the same key particle as before ($\sigma_{\rm K}=\sigma$).

Fig.~\ref{fig:LK-rattle-eff} shows results for $\Weff$ and $\Wapp$ in this case, which can be compared with Fig.~\ref{fig:LK-eff}.  Considering first $\Weff$, the general structure is very similar. The main notable features are that the bound region is more spread out but the depletion attraction is weaker ($\Weff$ less negative) due to the imperfect fit of the key within the lock.  There is also a region where the depletion interaction is repulsive, which is located near to the bound region.  This effect is due to layering of the (hard) depletant particles near the surface of the lock.

The method for inferring the approximate effective potential $\Wapp$ follows closely that of the previous section.  The main differences are as follows.  The specific binding regime is encapsulated by a single region with $d_c \leq r < d_c+q\sigma/3$, which plays the part of the innermost region $X_{{\rm LK},n=1}$ for exactly-fitting key.  The non-specific region is identical to that of the previous section.  The intermediate regime is treated in the same way as before (separated into two regions according to the separation and 180 regions according to angle), the only difference being that the function $r_0(\theta)$ which describes the excluded volume of the lock is slightly more complex.

Comparing Fig.~\ref{fig:LK-rattle-eff}(a,b), one sees that the region of repulsion between lock and key is not captured by this method.  Generalisation of the method to include this region would be straightforward, but the effect is relatively weak in this case so we have ignored it for the purposes of this study.  Also, it is apparent from Fig.~\ref{fig:LK-rattle-eff} that the strength of the depletion potential in the intermediate regime (as defined here) is comparable with its strength in the specifically bound regime, due to the more delocalised nature of the bound state.  

\subsection{Summary of results for effective interactions between lock and sphere particles}

We have demonstrated that the method of Sec.~\ref{sec:theory} can be used to describe the effective interactions between lock and key particles.  This method can be fully automated, so even if the parameterised effective interactions depend on a large number of parameters, these can be easily extracted from available simulation data.  However, two comments are in order.

First, the aim of the effective potential is to allow efficient simulation of a system of many interacting colloids, but any such application requires an effective interaction potential between the locks -- we discuss this interaction in the next section .  Second, for the simple systems considered so far, where the effective interactions depend on only two co-ordinates, one can imagine defining the effective potential by using large lookup tables based on the results in Figs.~\ref{fig:LK-eff}(a) and \ref{fig:LK-rattle-eff}(a), without defining regions associated with lock-and-key and specific binding.
However, for the lock-lock interactions described in the next section, the effective potential cannot be described by a simple two-dimensional histogram -- it depends on a set of four co-ordinates which are required in order to specify the relative position and orientation of the locks.  In that case a direct parameterisation of the effective potential would require a four-dimensional histogram instead of the two-dimensional histograms in Fig.~\ref{fig:LK-eff}(a).  This is impossible for practical purposes, so the theoretical approach described in Sec.~\ref{sec:theory} becomes essential. Indeed, we note that most previous studies have concentrated on interactions between locks and spheres~\cite{Odriozola:2008uq,Odriozola2013,Melendez2015,Villegas2016}, presumably because of the difficulty of characterising the lock-lock interaction.

\section{Results -- interaction between lock particles}
\label{sec:lock-lock}

In this section we consider effective interactions between indented colloids with $\sigma_{\rm c}=\sigma$ and $d_{\rm c}=0.5\sigma$: these are the same particles considered in Sec.~\ref{subsec:lk-exact}.  However, as far as possible, we describe our methods in a way that is easily generalized to other values of the lock shape parameters.

\subsection{Choice of co-ordinate system and analogy with lock-key binding}

\newcommand{\thetaR}{\theta_{\rm R}}
\newcommand{\thetaI}{\theta_{\rm I}}

The definition of an approximate effective potential $\Wapp$ for interacting lock particles requires a suitable co-ordinate system, which we take as in Fig.~\ref{fig:lock-lock-coord}.  Specifically, consider two particles, let the positions of their geometrical centres be $\bm{R}_{1},\bm{R}_{2}$ and their orientations be $\bm{n}_1,\bm{n}_2$. The distance between them is $r = |\bm{R}_1-\bm{R}_2|$.  We define three angles by $\cos\theta_1=\bm{n}_1\cdot (\bm{R}_2-\bm{R}_1)/r$, $\cos\theta_2=\bm{n}_2\cdot (\bm{R}_1-\bm{R}_2)/r$ and $\cos\phi=\bm{n}_1\cdot\bm{n}_2$, with all three angles chosen in the range $[0,\pi]$.  To ensure that the potential is symmetric under interchange of the two particles, it is useful to define a \emph{relevant} angle $\theta_{\rm R}=\min(\theta_1,\theta_2)$ and an \emph{irrelevant} angle $\theta_{\rm I}=\max(\theta_1,\theta_2)$.  The naming of the irrelevant angle anticipates the fact that our approximate effective interaction will not depend explicitly on $\thetaI$: see below.  On the other hand, the role of the relevant angle $\thetaR$ in this interaction is analogous to the role of the angle $\theta$ in the lock-sphere interaction considered in Sec.~\ref{sec:lock-key}.

To illustrate the analogy with the lock-sphere interaction, we define the probability density for $(r,\cos\theta_{\rm R})$, based on the data for the two particle system.  That is, 
\begin{multline}
p^\eta(\hat{r},\hat{c}_r) = \frac{1}{2Z_2^\eta} \int  \delta(\hat{r}-r) \delta(\hat{c}_r-\cos\theta_{\rm R}) 
\\ \times \ee^{-v_{\rm eff}(\bm{R}_1,\Omega_1,\bm{R}_2,\Omega_2)} \mathrm{d}\bm{R}_1 \mathrm{d}\Omega_1 \mathrm{d}\bm{R}_2 \mathrm{d}\Omega_2
\end{multline}
[The notation here is that $r=|\bm{R}_1-\bm{R}_2|$ is the particle separation (which is being integrated over) and $\hat{r}$ is the value of the separation at which the probability density is evaluated.  Similarly $\theta_{\rm R}$ is the relevant angle (which is being integrated) and $\hat{c}_{\rm r}$ is the argument of the probability density. In the following, we omit the hats in cases where this does not lead to any ambiguity.]

The distribution $p^\eta({r},{c}_r)$ can be estimated numerically by binning and histogramming data for $r$ and $\cos\theta_{\rm R}$  from a computer simulation of two colloidal particles, interacting with a depletant.
Then we define a free energy (strictly, a free energy {difference}) that depends on these two co-ordinates, as
\begin{equation}
w^\eta(r,c_r) = -\log \left[ \frac{p^\eta(r,c_r)}{p^0(r,c_r)} \cdot \frac{Z_2^\eta}{Z_2^0} \right]
\end{equation}
where $p^0$ indicates $p^{\eta=0}$, as usual.  The partition functions $Z_2^{\eta,0}$ are evaluated using (\ref{equ:B2-Z-PR}). 
For the lock-key system, this free energy is equal to the effective interaction.   In the lock-lock system, this is not the case because the effective interaction $\Weff$ depends on two additional variables $\phi,\theta_{\rm I}$, as we discuss below.  However, $w(r,c_r)$ is a useful quantity to measure, because it shows the values of $r$ and $\thetaR$ which are enhanced or suppressed by the depletant.

\begin{figure}
\includegraphics[width=8.5cm]{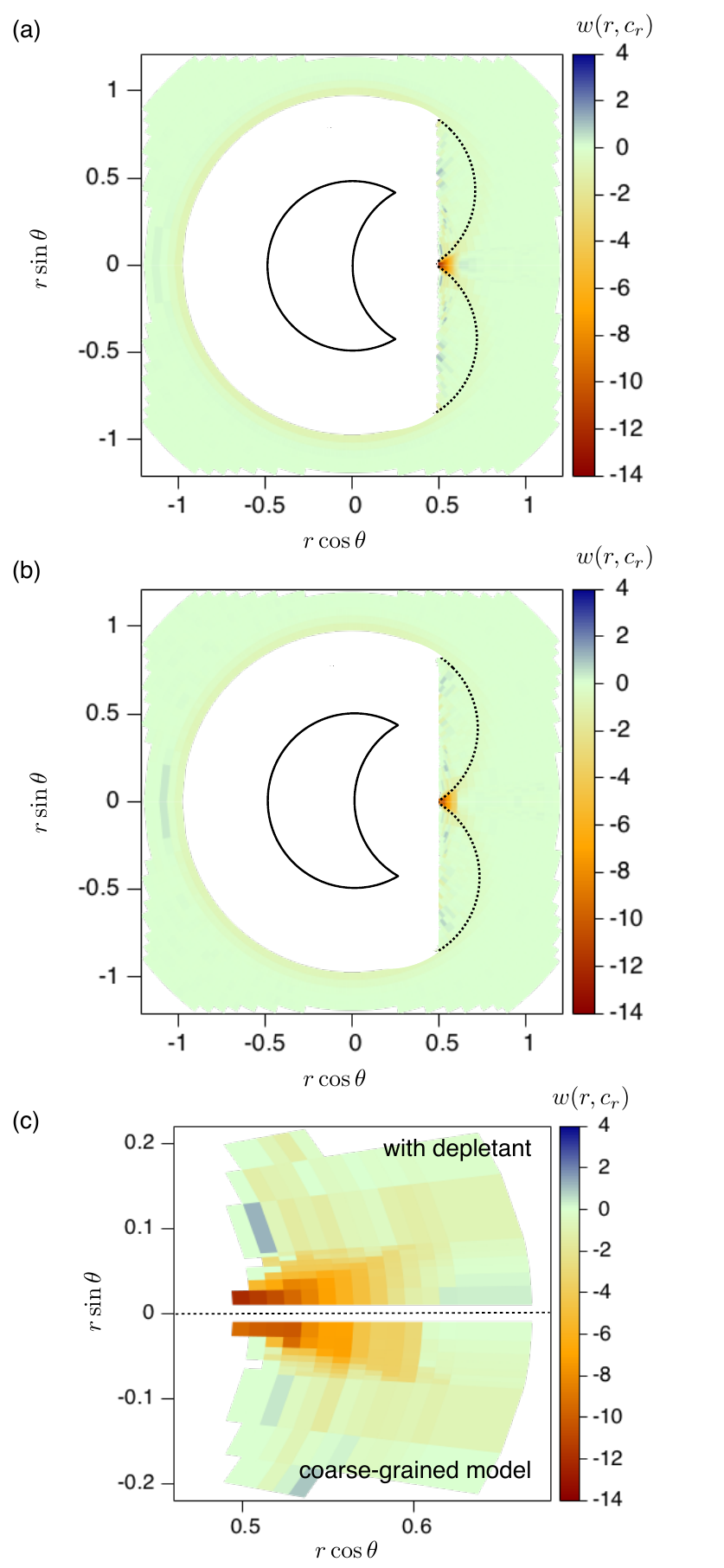}
\caption{(a) Free energy $w(r,c_r)=w(r,\cos\theta)$ evaluated for two locks interacting with a depletant.  Comparing with Fig.~\ref{fig:LK-eff}, one sees regions associated with specific and non-specific binding, as expected.  The dotted line is Eq.~(\ref{equ:r0-overlap}) which is the distance of closest approach of a spherical key particle.  In contrast to the lock-key case, there is a finite probability of the two particles approaching closer than this boundary: this corresponds to lock particles approaching each other in a ``mouth-to-mouth'' configuration. %(b, to be REVISED) approximate (coarse-grained) effective interaction. 
(b)~Free energy $w(r,c_r)$ calculated for locks interacting via the effective potential (without depletant).
(c)~Enlarged figure showing the free energy in the lock-key bound region, where the system with depletant (data from panel (a)) is compared with the coarse-grained (approximate) potential (data from panel (b)).
}
\label{fig:ll-eff-r-cr}
\end{figure}

The free energy $w(r,c_r)$ is plotted in Fig.~\ref{fig:ll-eff-r-cr}(a). This shows that the effective potential is attractive in two main regions, which correspond to the specific (lock-and-key) and non-specific (back-to-back) binding modes shown in Fig.~\ref{fig:lock-lock-coord}(b,c).  Comparing with Fig.~\ref{fig:LK-eff}, there is also finite probability density for the lock particles to approach each other more closely than is possible for a lock and key.  The effect arises when the two lock mouths (indentations) are oriented towards each other, as in Fig.~\ref{fig:lock-lock-coord}(d).  In that case one sees that a particle overlap would result if either lock was replaced by a sphere.

\subsection{Choice of regions $X$}

As for the case of a sphere interacting with a lock, we define the approximate effective potential $\Wapp$ by dividing the parameter space into regions and considering them in turn.
Given the similarities between Fig.~\ref{fig:ll-eff-r-cr}a and Fig.~\ref{fig:LK-eff}a, the natural choice is to retain the three main regimes shown in Fig.~\ref{fig:LK-schem}, associated with specific (lock-and-key) binding, non-specific (back-to-back) binding and an intermediate regime.  In addition, there is a fourth regime which consists of those values of $r,c_r$ which were inaccessible in the lock-sphere case: this corresponds to the mouth-to-mouth case in Fig.~\ref{fig:lock-lock-coord}(d).  

In the following, we consider these four regimes in turn.  Each regime is defined by constraints on the values of $r$ and $c_r=\cos\theta$.  Within each regime, we identify different regions and we either use (\ref{equ:Wapp-P}) to assign a value of $\Wapp$, or in some cases we use an alternative but similar method (see below).  The resulting approximate effective interaction is independent of $\thetaI$.   In the back-to-back (non-specific) regime and the intermediate regime it is also independent of $\phi$, and the effective interaction is very similar to the lock-sphere interaction.  In the lock-and-key regime and the mouth-to-mouth regime, the effective potential does depend on the angle $\phi$. 

\subsubsection{Specific (lock-and-key) binding}
\label{subsec:ll-specific}

\begin{figure}
\includegraphics[width=7cm]{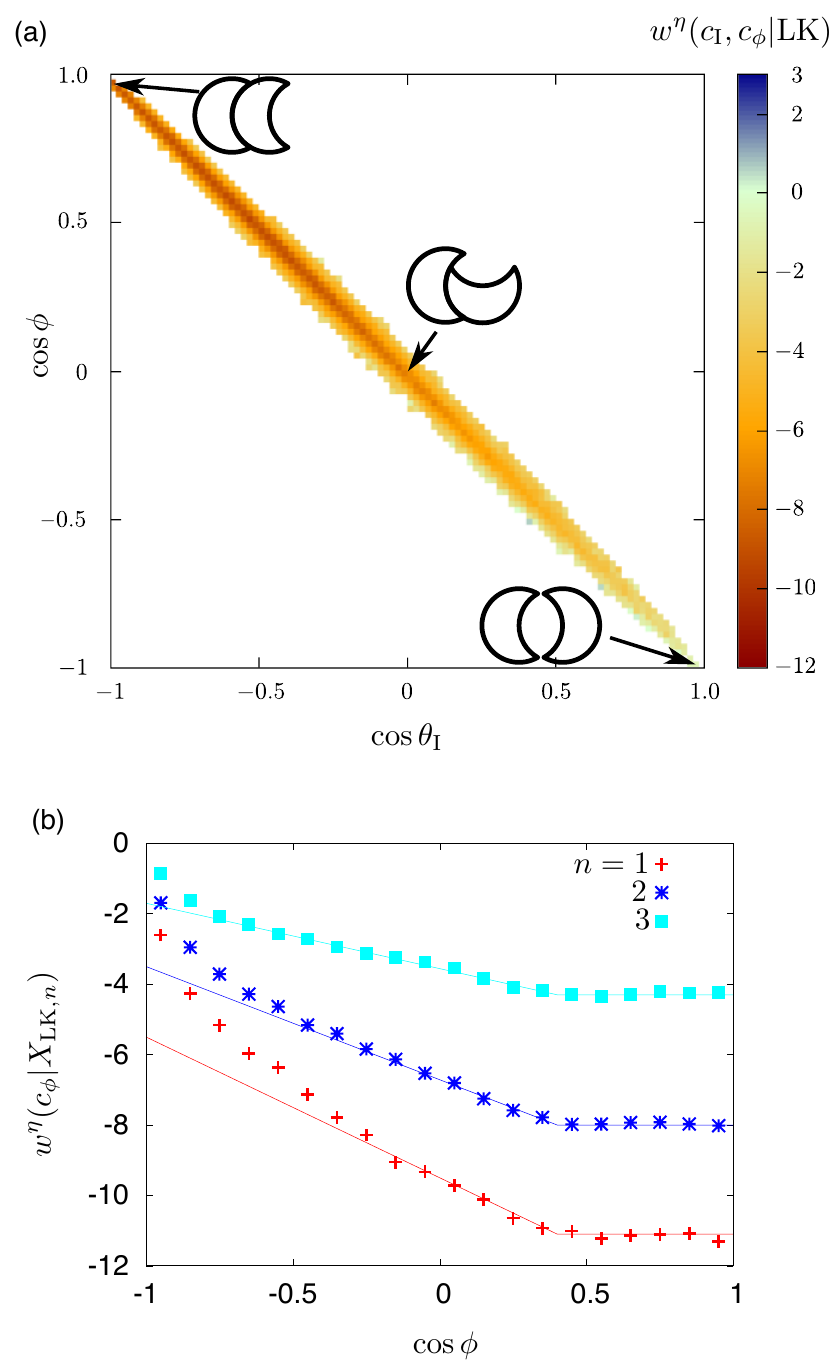}
\caption{(a) Free energy as a function of $(c_{\rm I},c_\phi)=(\cos\thetaI,\cos\phi)$, given that the particles are in the lock-and-key binding regime.  The attraction is strongest when $\phi$ is small ($c_\phi\approx 1$).
(b) Analogous free energy as a function of $c_\phi$ only, but subdivided into three regions according to the distance of between the lock particles.  The points show the measured free energy and the lines show the $\phi$-dependence of the approximate coarse-grained potential $\Wapp$.  The angular dependence of $\Wapp$ does not match perfectly but the contributions of the $\phi$-dependent regions to the second virial coefficient do match the true effective potential. }
\label{fig:ll-ci-cphi}
\end{figure}

The specific (lock-and-key) binding regime in Fig.~\ref{fig:LK-schem} has an analogue for the interacting lock case which is
\begin{equation}
r_0(\theta) < r < d_c+q\sigma
\label{equ:ineq-A}
\end{equation}
This region corresponds to the bound area in Fig.~\ref{fig:LK-eff}.
 The explicit lower bound on $r$ comes from (\ref{equ:r0-overlap}) and means that this specific binding regime only includes states where one of the locks could be replaced by a spherical key particle without overlapping the other lock.  (Particles approaching more closely than this will be considered in the mouth-to-mouth regime, see below.)

Within this regime, the effective interaction depends strongly on the angle $\phi$.  
To quantify this, define an (un-normalised) probability density for the cosines of the angles $\theta_{\rm I},\phi$, 
for states restricted to this lock-and-key binding regime
\begin{multline}
p^\eta(\hat{c}_i,\hat{c}_\phi|{\rm LK}) = \frac{1}{2Z_2^\eta} \int_{\rm LK} \delta(\hat{c}_i-\cos\theta_{\rm I}) \delta(\hat{c}_\phi-\cos\phi)
\\ \times \ee^{-v_{\rm eff}(\bm{R}_1,\Omega_1,\bm{R}_2,\Omega_2)} \mathrm{d}\bm{R}_1 \mathrm{d}\Omega_1 \mathrm{d}\bm{R}_2 \mathrm{d}\Omega_2
\label{equ:pcicp}
\end{multline}
where the integral domain is given by (\ref{equ:ineq-A}).
There is an associated free energy
\begin{equation}
w^\eta(c_i,c_\phi|{\rm LK}) = -\log \left[ \frac{p^\eta(c_i,c_\phi|{\rm LK})}{p^0(c_i,c_\phi|{\rm LK})} \cdot \frac{Z_2^\eta}{Z_2^0} \right]
\end{equation}
which is negative if the depletant enhances the probability of finding a particular value of these co-ordinates, given that (\ref{equ:ineq-A}) is satisfied.  This free energy is plotted in Fig.~\ref{fig:ll-ci-cphi}(a).  Two points are noteworthy: first, all the data lie close to a diagonal line in this two-dimensional space.  The reason is purely geometrical -- if one fixes $\theta_R=0$ then one must have $\phi=\pi-\theta_{\rm I}$, just from the definition of the co-ordinate system and independent of the colloid shapes.  Given that all data in Fig.~\ref{fig:ll-ci-cphi} come from the specific binding regime and therefore have small values of $\theta_{\rm R}$, this explains the inaccessible (white) regions in Fig.~\ref{fig:ll-ci-cphi}.  Second, there is significant dependence of the effective potential on these angles, with strong interactions when $\phi$ is small ($c_\phi\approx 1$) and $\theta_{\rm I}$ is large ($c_i\approx -1$).  The reason for this strong dependence is illustrated by the snapshots in the same figure, which show that only when $c_\phi$ is large does one observe strong lock-and-key binding.

Motivated by the one-dimensional structure in Fig.~\ref{fig:ll-ci-cphi}(a), we define an analogous free energy for $c_\phi$ alone.
\begin{equation}
w^\eta(c_\phi|X) = -\log \left[ \frac{p^\eta(c_\phi|X)}{p^0(c_\phi|X)} \cdot \frac{Z_2^\eta}{Z_2^0} \right]
\end{equation}
as a function of the single variable $c_\phi$, now restricted to a specific region $X$.  We subdivide the lock-and-key binding regime into three regions $X_{{\rm LK},n}$, according to the same distance cutoffs used for the analogous case in (\ref{equ:lk-bound-ineq}).  Fig.~\ref{fig:ll-ci-cphi}(b) shows the effective interactions as a function of $\phi$ for these three regions.

The $\phi$-dependence of $\Weff$ could be captured approximately with a piecewise constant function but we choose an alternative approach here.  For $\cos\phi \geq \cos\phi^*$ and within each region $X_{{\rm LK},n}$, 
we use a constant value of $\Wapp$.  We take $\cos\phi^*=0.4$, consistent with Fig.~\ref{fig:ll-ci-cphi} and $\Wapp$ is determined from (\ref{equ:Wapp-P}) using the regions $X_{{\rm LKf},n}$ obtained from $X_{{\rm LK},n}$ by restricting also to $\cos\phi\geq\cos\phi^*$.  

For $-1\leq\cos\phi<\cos\phi^*$ we take $\Wapp$ to have linear dependence on $\cos\phi$, to reflect the structure in $w^\eta(c_\phi|X)$.  The result is shown in Fig.~\ref{fig:ll-ci-cphi}(b).  In the linear regime, we fix $\Wapp$ to be continuous at $\phi^*$ and we choose the value of the intercept $\Wapp(c_\phi=-1)$ so that the contribution of region $X_{{\rm LK},n}$ to the second virial coefficient matches the corresponding value of the exact effective interaction: see Appendix~\ref{app:wapp-fit}.

One sees from Fig~\ref{fig:ll-ci-cphi} that this method overestimates the strength of the effective interaction for negative values of $c_\phi$, but this is not a serious approximation since such configurations are rather rare in any case.  The poor agreement for $c_\phi\approx-1$ occurs because the second virial coefficient is dominated by values of $\phi$ for which the interaction is strong: in that case the factor ${\rm e}^{-\beta v_{\rm eff}}$ in (\ref{equ:b2}) is large.  Hence, the method of matching second virial coefficients tends to parameterise the effective interaction most accurately in regions of strong binding (in this case, large $c_\phi$).  In fact, this feature is a key strength of the method, since regions of strong binding are the most important feature that the coarse-grained model should capture.  \rlj{Note that an alternative strategy might have been to fit the \emph{values} of the effective potential in Fig.~\ref{fig:ll-ci-cphi}(b) instead of matching $B_2(X)$.  This would lead to a better apparent fit in the figure, but the physical behaviour of the system is controlled by $B_2(X)$, so we would expect the resulting model to be less accurate in predicting this physical behavior.}
(One might also improve the approximate effective interaction shown in Fig.~\ref{fig:ll-ci-cphi}, for example by following the parameterisation strategy used in the intermediate regime for the lock-sphere interaction.  The difficulty in this case is that the function $w^\eta(c_\phi)$ is rather expensive to estimate numerically, since such configurations are very rare in the simulations of locks without depletant.  Hence our use of a piecewise linear approximation.)

As a final test of our effective potential in this regime, we consider Fig.~\ref{fig:ll-eff-r-cr}(c), which is an enlarged plot of $w^\eta(r,c_r)$, concentrating on lock-and-key binding.  The true free energy is shown in the upper panel, and is compared with the same free energy evaluated using the approximate effective interaction $\Wapp$.  The agreement is good.  

\subsubsection{Intermediate and back-to-back regimes}

The effective potential in the back-to-back (non-specific) regime and the intermediate regime follows exactly the procedure described in Sec.~\ref{subsec:lk-exact}, except that the definition of the intermediate regime includes a constraint that $r>r_0(\theta)$, as in (\ref{equ:ineq-A}).  Since the regions are defined in this way, $\Wapp$ does not depend on $\theta_{\rm I},\phi$ within these areas.  Of course this represents an approximation (particularly in the intermediate regime) but the probability of binding in that regime is not high so this approximation does not have a strong impact on the resulting coarse-grained model.

\begin{figure}
\includegraphics[width=7cm]{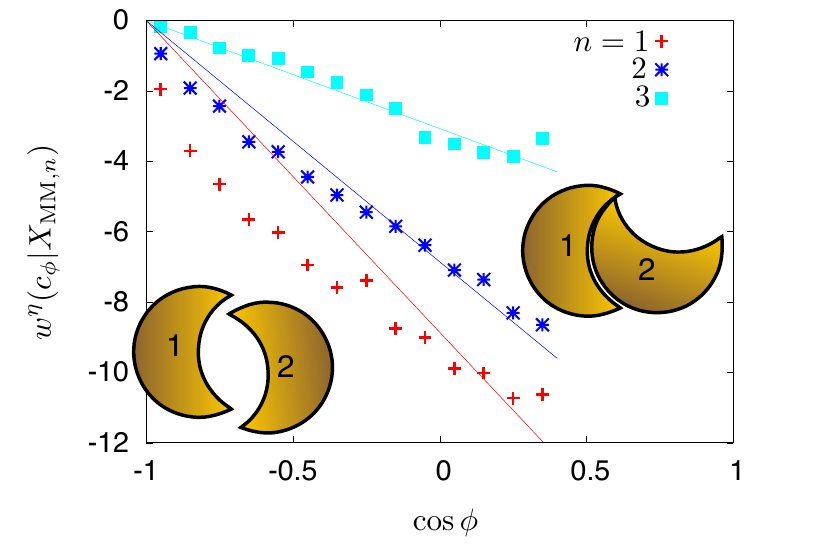}
\caption{Free energy $w^\eta(c_\phi|X_{{\rm MM},n})$ for mouth-to-mouth regions, analogous to the bound case shown in Fig.~\ref{fig:ll-ci-cphi}b.  Points show the free energy evaluated from simulation and solid lines show the effective potential $\Wapp$.  Configurations with high and low values of $\cos\phi$ are illustated. The difference between this case and Fig.~\ref{fig:ll-ci-cphi} is that if the locks labelled 2 in these configurations were replaced spheres then they would overlap with the locks labelled 1: this is the distinction between the mouth-to-mouth and lock-and-key binding regimes.}
\label{fig:ll_mm_cp}
\end{figure}

\subsubsection{Mouth-to-mouth regime}

The mouth-to-mouth binding regime is defined by those values of $(r,\cos\thetaR)$ that would not be possible for a lock interacting with a spherical particle, that is 
\begin{equation}
r<r_0(\theta)
\label{equ:reg-mm}
\end{equation}
with $r_0(\theta)$ given by (\ref{equ:r0-overlap}).  These are binding modes which are impossible for a spherical key interacting with a lock, but are possible for two locks.  The resulting effective interactions are strong only for small $r$ so within the mouth-to-mouth regime we take $\Wapp\neq0$ only for
\begin{equation}
r < d_{\rm c} + q\sigma
\label{equ:mm-cut}
\end{equation}
which is the same large-$r$ cutoff as we used for the specific (lock-and-key) binding regime.  Insisting always that (\ref{equ:reg-mm}) holds we then define three regions $X_{{\rm MM},n}$ (with $n=1,2,3$) using the same distance cutoffs (\ref{equ:lk-bound-ineq}) as in the lock-and-key (specific) binding regime.

The primary contribution to the resulting effective interaction arises from configurations that are very close to the bound regime.  The generalisation of Fig.~\ref{fig:ll-ci-cphi}(b) for these regions is shown in Fig.~\ref{fig:ll_mm_cp}.  Compared to the lock-and-key binding regime, the accessible range of $c_\phi$ is reduced, due to the excluded volume interactions.  For the $c_\phi$ values that are accessible, we choose a linear dependence of the effective potential on $\phi$, as in the lock-and-key binding regime.  We fix the potential to zero at $c_\phi=-1$ and we adjust the slope of the effective potential using the method described in Appendix~\ref{app:wapp-fit}, so that the contribution of this region to the second virial coefficient matches between exact and approximate effective potentials.  The potentials themselves do not agree exactly but this has little effect on the overall physics because the probability of binding in this regime is much smaller than that of regular lock-and-key binding.  [In fact, the entropy associated with binding in this regime is very low, due to the strong geometrical constraints on ($r,\thetaR,\phi)$.]  

Finally, we note that the good agreement between the fully interacting model and the coarse-grained model in Fig.~\ref{fig:ll-eff-r-cr}(c) does depend on a suitable $\phi$-dependent parameterisation of $\Wapp$ in the mouth-to-mouth regime, since the range of accessible values of $\phi$ depends strongly on $r,c_r$.  In particular, using the regions $X_{{\rm MM},n}$ but neglecting the $\phi$-dependence of $\Wapp$ in this regime leads to $w(r,c_r)$ depending only on $r$, which is not consistent with the true free energy.

\begin{figure}
\includegraphics[width=7cm]{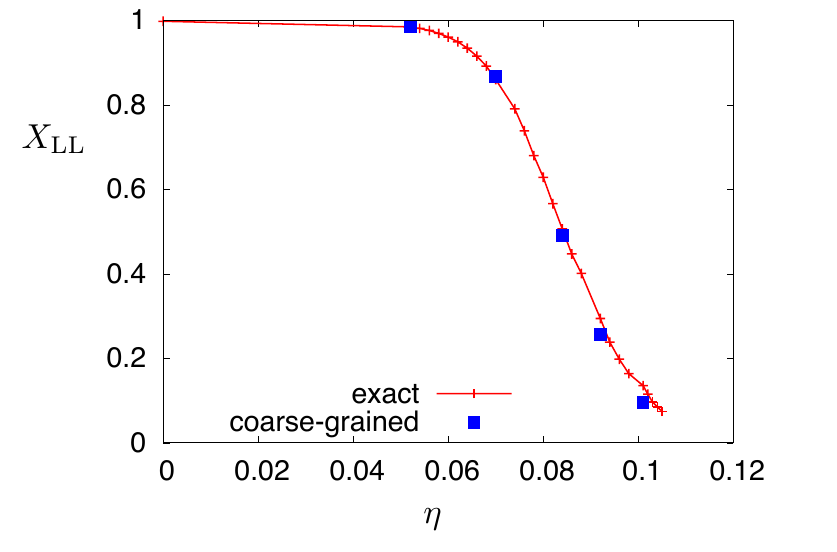}
\caption{Fraction $X_{\rm LL}$ of unoccupied lock-and-key binding sites in a system of indented colloids, as a function of the depletant volume fraction.  The ``exact'' data are taken from Ref.~\onlinecite{Ashton2013}, and involve expensive simulation of lock particles interacting with a depletant.  The ``coarse-grained'' data were obtained from a (much shorter) simulation of colloids interacting by the effective potential described in this work.  The simulations involve $N=60$ colloids at number density $\rho=0.2\sigma^{-3}$}
\label{fig:ll_chain}
\end{figure}

\subsection{Verification}

With these effective potentials in place, it is straightforward to simulate systems of indented colloids, interacting by the approximate effective potential $\Wapp$.  Fig.~\ref{fig:ll_chain} shows results for this case, compared to results for the fully interacting system of colloids and depletant, from Ref.~\onlinecite{Ashton2013}.  As in that work, it is convenient to measure the number of lock-and-key bonds $N_{\rm LL}$ between the colloids, and to normalize this by the total number of particles, $N$.  Then we define $X_{\rm LL} = 1 - (N_{\rm LL}/N)$ which is the fraction of colloidal indentations that are not involved in any lock-key bond.  Hence $X_{\rm LL}$ decreases from a value close to unity at $\eta=0$ to a value close to zero when the interactions are very strong.  

The agreement in Fig.~\ref{fig:ll_chain} between the exact and coarse-grained models is good.  Deviations are visible for large $\eta$: we note that in this case, equilibration of simulations with depletant is challenging, and it is possible that the deviations between exact and coarse-grained models are due to a failure to equilibrate the fully-interacting system.  We also note that theoretical predictions of $X_{\rm LL}$ can be obtained from the contribution to the second virial coefficient from lock-and-key bonding, using Wertheim's theory\cite{Ashton2013}, so the fact that the coarse-grained interaction matches this contribution means that agreement between exact and coarse-grained models should be expected.  However, the agreement of the coarse-grained model with the exact results in Fig.~\ref{fig:ll_chain} is significantly better than the agreement with Wertheim's theory in Ref.~\onlinecite{Ashton2013}, showing that it is not sufficient just to match this second virial coefficient: a reasonably accurate description of  effective interaction is also required in order to achieve this agreement.

\section{Simplified lock-lock potential}
\label{sec:simp}

\newcommand{\Weps}{W_{\epsilon}}

Finally, we discuss the connection of the results of this work to the effective potential used in Ref.~\onlinecite{Ashton2015-porous}, which was developed with the aid of some of the results presented here.  We refer to that interaction potential as $\Weps$ to avoid confusion with the coarse-grained potential $\Wapp$ discussed here.

\subsection{Comparison with results for colloidal polymers}

The potential $\Weps$ is defined in Ref.~\onlinecite{Ashton2015-porous}: we give a brief recap here.
In the back-to-back regime, $\Weps=-\epsilon_{\rm BB}$ throughout a region defined by $\sigma<r<\sigma(1+\xi)$ and $\theta_{\rm R}>\theta^*$.  The value of $\xi$ is fixed at $0.1$.  \rlj{In terms of Fig.~\ref{fig:LK-schem}, this corresponds to using a single region for the whole non-specific regime, instead of two regions as in this work.}

\rlj{In the intermediate regime identified in this work (recall Fig.~\ref{fig:LK-schem}), we take $\Weps=0$, for simplicity.  (The justification for this assumption is that the intermediate binding regime is rarely seen in practice since it competes with specific lock-and-key binding, which is much stronger.)}

For the specific (lock-and-key) regime, $\Weps$ is defined in terms of a single region defined as $r<d_c+\sigma\xi$, which includes both the specific binding and mouth-to-mouth binding regimes described here for the lock-lock interaction. 
\rlj{Comparing with Fig.~\ref{fig:LK-schem}, the three regions in the ``specific binding'' regime are replaced by a single region, which also includes the ``mouth-to-mouth'' regime (not shown in Fig.~\ref{fig:LK-schem}).
Within this regime $\Weps$ depends on $\phi$, in a similar way to $\Wapp$, except for two simplications.  First, the linear segment of the effective potential shown in Fig.~\ref{fig:ll-ci-cphi}b is constrained to reach zero when $\cos\phi=-1$; second, the linear segment of the effective potential in Fig.~\ref{fig:ll_mm_cp} is taken exactly to equal to that in Fig.~\ref{fig:ll-ci-cphi}b.  From inspection of Figs.~\ref{fig:ll-ci-cphi},\ref{fig:ll_mm_cp}, these additional constraints will lead to a slightly less accurate representation of the exact data.  However, the key advantage is that when inferring the effective potential $\Weps$ from data for colloids interacting with depletant, the value of $W_\epsilon$ at $\cos\phi=+1$ is a single adjustable parameter that is chosen to match $B_2(X)$ for this region.  This is much simpler than inferring $\Wapp$, which requires six parameters for the bound regime and three more for the mouth-mouth regime.
}
%one has $\Weps=-\epsilon_{\rm LK}$ for $\cos\phi > \cos\phi^*$, while for angles $\phi>\phi^*$ (that is, $\cos\phi<\cos\phi^*$) the effective potential depends linearly on $\cos\phi$ (with $\Weps=0$ at $\cos\phi=-1$ and $\Weps=-\epsilon_{\rm LK}$ at $\cos\phi=\cos\phi^*$).  

%Comparing $\Weps$ with the potential $\Wapp$ considered here, both potentials treat the back-to-back regime similarly, except $\Weps$ classifies the separation $r$ using just one region instead of two.  The lock-and-key binding regime is also similar, except $\Weps$ again classifies the separation $r$ using just one region (instead of three); also $\Weps\to0$ as $\phi\to-1$ instead of the non-zero intercept used in Fig.~\ref{fig:ll-ci-cphi}b.  In the mouth-to-mouth regime $\Weps$ has the same $\phi$-dependence as in the lock-and-key regime: that is, the differences between Fig.~\ref{fig:ll-ci-cphi}b and Fig.~\ref{fig:ll_mm_cp} are neglected, leading to a simpler description than that used in $\Wapp$.  Also, $\Weps=0$ throughout the intermediate regime which again leads to a simpler description than that used in $\Wapp$.

\begin{figure}
\includegraphics[width=8cm]{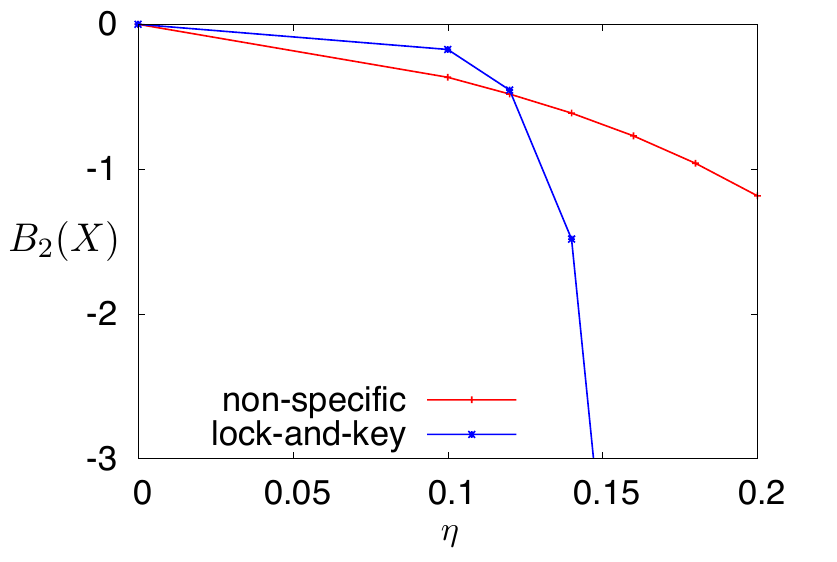}
\caption{\rlj{Data showing contributions to the second virial coefficient for specific (lock-key) and non-specific (back-to-back) binding, for a system with AO depletant with size ratio $q=0.126$, as used in Ref.~\onlinecite{Ashton2015-porous}.  The behaviour is very similar to Fig.~\ref{fig:lk-b2}, although the larger $q$-value means that relative strengths of specific and non-specific interactions are more similar and the overall strength of the interaction is lower (at fixed volume fraction $\eta$).}}
\label{fig:ao-well}
\end{figure}

\rlj{As a result of these simplifications, $\Weps$ depends on just three parameters: the range $\xi$ and the interaction strengths for specific and non-specific binding.  Fixing $\xi=0.1$ and given a depletant at volume fraction $\eta$ with size ratio $q$, one may derive an effective potential $\Weps$ by matching contributions to the second virial coefficient from the lock-and-key and back-to-back regions.  
The relevant contributions to the second virial coefficient are shown in Fig.~\ref{fig:ao-well}, for an AO depletant with size ratio $q=0.126$ (see also the Supplemental Material of Ref.~\onlinecite{Ashton2015-porous}).  Fig.~\ref{fig:ao-well} is qualitatively similar to the results for a lock-sphere system shown in Fig.~\ref{fig:lk-b2} of this work, although we note that Fig.~\ref{fig:ao-well} shows results for a \emph{lock-lock} interaction mediated by an \emph{ideal} (Asakura-Oosawa~\cite{Asakura1954}, AO) depletant.  (As expected, the larger depletant used in Fig.~\ref{fig:ao-well} leads to weaker interactions compared to Fig.~\ref{fig:lk-b2}, when comparing at fixed volume fraction $\eta$.  It also leads to a smaller difference in strength between specific and non-specific binding modes).}

\rlj{The argument in Ref.~\onlinecite{Ashton2015-porous} is that $\Weps$ provides a semi-quantitative model of lock-and-key colloids interacting with a depletant, and that matching the two values $B_2(X)$ shown in Fig.~\ref{fig:ao-well} allows the behaviour of a range of lock-and-key systems to be modelled using a single effective potential.  Using this model revealed novel phase behaviour, including porous liquid phases~\cite{Ashton2015-porous}, with similarities to those found in patchy-particle models\cite{Bianchi2006,Bianchi2008,Ruzicka:2011ud}.  In order to obtain accurate results for a \emph{specific} microscopic model such as the hard-sphere depletant used in this work, we expect that using the more complicated potential $\Wapp$ would yield a more accurate match with the underlying microscopic model, but we would expect the observed phase behaviour to be robust, especially given the theoretical predictions (based on Wertheim's theory~\cite{Wertheim:1984zr}) that this behavior is controlled by the second virial coefficients for specific/non-specific binding~\cite{Ashton2015-porous}.}

\rlj{
\subsection{Comparison between hard sphere and ideal depletant}

We noted above that the behavior discussed in Ref.~\onlinecite{Ashton2015-porous} is based on a model with an ideal (AO) depletant.  In that case, the (spherical) depletant particles describe polymer chains so they can overlap with each other, although they cannot overlap with the colloidal particles.  This feature means that the strength of the AO interaction can be expressed geometrically in terms of the overlap between geometrical shapes.  However, in contrast to the simple situation of spheres interacting with each other or with walls, analytic calculations of AO interactions between indented colloids are limited to idealised geometries~\cite{Ashton2015-wall}, although numerically exact calculations of lock-key interactions have been performed~\cite{Villegas2016}.  For the full lock-lock interaction considered here, one encounters the same difficulties in the AO case as for the hard sphere depletant -- the effective interaction is a function of four variables and requires an approximate representation.

We already showed results in Fig.~\ref{fig:ao-well} for the second virial coefficients based on an AO depletant, which reveal similar qualitative behaviour to Fig.~\ref{fig:lk-b2}.  If we compare the effective potentials for AO and hard sphere depletants in more detail,
we find the behavior is very similar.  Since the AO interaction is not analytically tractable in this system, a full analysis of these differences would require the calculation shown here to be repeated for the AO case, which is beyond the scope of this paper.  However, we note that the main difference between AO and hard-sphere depletant is the layering effect of the hard-sphere depletant near the colloid surfaces, which leads to the repulsive interactions between colloids that are apparent for intermediate separations in Fig.~\ref{fig:LK-rattle-eff}.  A similar effect is present in Figs.~\ref{fig:LK-eff}a and \ref{fig:ll-eff-r-cr}a but is weaker in those cases and hence not so visible in the plots.  In all cases, this effect has been neglected in the effective potential, for simplicity.  In this sense, the effective potential that we describe in this work is also a rather accurate model for a system with an ideal depletant.  
}

\section{Conclusions}
\label{sec:conc}

We have presented a general technique for obtaining
coarse-grained effective potentials
that approximate the interactions between anisotropic colloids, immersed in a
depletant. The method takes data for the joint probability
distribution of the relative positions and orientations of a pair
of colloids, obtained from simulations that include the depletant explicilty.  
These data were obtained in this case by the geometrical
cluster algorithm, although other methods can also be used. These data are then used to derive an
approximate depletion potential which is a piecewise-constant function of the relative positions and orientations of the colloids.  Deriving this potential requires a decomposition of the two-particle configuration space, which is chosen according to physical reasoning.  This decomposition fully specifies the effective potential, which can then be inferred automatically from the simulation data, by matching the
second virial coeffients between the full and approximate effective
potentials in each region.  This method thereby ensures thermodynamic consistency
at the level of the free energy of bonding.

We have illustrated our approach for lock and key colloids, showing
how one decomposes the domain of position and orientation into
appropriate regions and implements the matching strategy within each.
The resulting depletion potentials are not simple because describing
the relative position and orientation of anisotropic particles
requires several co-ordinates (for example, the interaction potential
between uniaxial particles depends on one distance and three angles).
Nevertheless, the accurate effective potentials that we derive allow
quantitative agreement with fully interacting systems of many
colloids. This was tested via a comparison of the self assembly of
indented colloids into chains, for which the depletion potential gave
excellent agreement with a GCA simulation of the full systems of hard
indented colloids plus depletant (Fig.~\ref{fig:ll_chain}), but at a
fraction of the computational cost.

There are many instances in which colloidal anisotropy is expected to
lead to interesting self assembly behaviour controlled by
depletion. By its entropic nature, the interaction is strongest
between surfaces with complementary shapes. However the overall scale
of the depletion interaction and the ratio of the strength of specific
to non-specific interactions can be controlled by changing the volume
fraction of depletant and the depletion-colloid size ratio. This makes
depletion a versatile interaction for the control of self
assembly. With the increasing ability to create colloids with a wide
variety of shapes, it is becoming practical to use depletion to
assemble these building blocks into designer structures~\cite{Anders:2014la,Ashton2015-porous,Rossi:2015aa}. If simulation
is to keep up with these advances and provide predictions as to the
types of assembled structures that might occur, it will be necessary
to have reliable coarse-graining strategies for describing the
effective interactions. Our method should be of use here and it would
be interesting to apply it to other anisotropic colloids\cite{Rossi:2015aa,Rossi:2011qd,Sacanna:2013aa,Mukhija:2011aa,Dugyala:2013aa}.

\begin{acknowledgments}
We thank the Engineering and Physical Sciences Research Council (EPSRC) for funding through grant EP/I036192/1.
\end{acknowledgments}

\begin{appendix}

\section{Deriving potentials $\Wapp$ that are not piecewise constant}
\label{app:wapp-fit}

In Sec.~\ref{subsec:ll-specific}, we explained that the specific interaction between lock particles is described by a piecewise linear effective potential.  Since this potential is not piecewise constant, its values cannot be inferred using (\ref{equ:Wapp-P}) and so a different method is required.  To explain this procedure in a general way, we consider an effective potential $\Wapp$ that depends on just one co-ordinate (in the case of Fig.~\ref{fig:ll-ci-cphi}b, this co-ordinate is $c_\phi$) and we describe the effective potential by a parameter $y$ (in this case $y$ is the intercept of the effective potential at $c_\phi=-1$).  Our aim is to find the value of $y$ such that $B_2^{\rm app}(X)$ defined in (\ref{equ:B2app}) matches the second virial contribution $B_2^\eta(X)$ defined in (\ref{equ:B2X}).

Recall that $p^\eta(c_\phi|X)$ is the unnormalised distribution of $c_\phi$ within region $X$, defined by analogy with (\ref{equ:pcicp}).  (Note that this distribution depends on the system size $V$ through the partition function $Z_2^\eta$.)  It is convenient to write
\begin{align}
B_2^\eta(X) & = \tfrac12 \int_X (1-\ee^{-\beta v_{\rm eff}} ) \mathrm{d}\bm{R}' \mathrm{d}\Omega'
\nonumber \\
& =  \tfrac12 V(X) - \tfrac{1}{2V} \int_X \ee^{-\beta v_{\rm eff}} \mathrm{d}\bm{R}\, \mathrm{d}\Omega\,  \mathrm{d}\bm{R}' \mathrm{d}\Omega' 
\nonumber \\ 
& = \tfrac12 V(X) - (Z_2^\eta/V) \int p^\eta(c_\phi|X) \mathrm{d}\phi
\nonumber \\ 
& = \tfrac12 V(X) - Z_2^\eta P_2^\eta(X) / V
\end{align}
Similarly
\begin{align}
B_2^{\rm app}(X) & = \tfrac12 \int_X (1-\ee^{-\beta v_{\rm app}} ) \mathrm{d}\bm{R}' \mathrm{d}\Omega'
\nonumber \\
& =  \tfrac12 V(X) - \tfrac{1}{2V} \int_X \ee^{-\beta v_{\rm app}} \mathrm{d}\bm{R}\, \mathrm{d}\Omega\,  \mathrm{d}\bm{R}' \mathrm{d}\Omega' 
\nonumber \\ 
& = \tfrac12 V(X) - (Z_2^0/V) \int_X p^0(c_\phi|X) \ee^{-W_{\rm app}(c_\phi|X)} \mathrm{d}c_\phi
\end{align}
where $W_{\rm app}(c_\phi|X)$ is the ($c_\phi$-dependent) approximate effective potential in region $X$ (which depends on the parameter $y$).

Noting from (\ref{equ:P2B2}) that  $P_2^0(X)=\int p^0(c_\phi|X)  \mathrm{d}c_\phi$, 
we define a normalised probability distribution for $c_\phi$ (within region $X$) as
$\tilde p^0(c_\phi|X) = p^0(c_\phi|X) / P_2^0(X)$.  
Finally, enforcing the constraint
that $B_2^\eta(X)=B_2^{\rm app}(X)$ we obtain 
\begin{equation}
 \int_X {\tilde p}^0(c_\phi|X) \ee^{-W_{\rm app}(c_\phi|X)} \mathrm{d}\phi =  \frac{P_2^\eta(X)}{P_2^0(X) } \cdot \frac{Z_2^\eta}{Z_0^\eta} .
\end{equation}
The right hand side of this equation is the same quantity that appears in (\ref{equ:Wapp-P}) and can be calculated from simulation data.  Calculation of the left-hand side from simulation data requires sufficient data to estimate (as a histogram) the distribution $\tilde p(c_\phi|X)$.  With this data in hand, the left hand side can then be calculated as an average with respect to this distribution, which depends on the parameter $y$.  A simple search over values of the parameter $y$ then yields an effective potential for which $B_2^\eta(X)=B_2^{\rm app}(X)$. 

\end{appendix}

\bibliography{extra}

\end{document}